\documentclass[]{aa} 

\usepackage{graphicx}
\usepackage{txfonts}
\usepackage[table]{xcolor}
\usepackage{amsmath}
\usepackage{amsfonts}
\usepackage{natbib}
\usepackage{longtable}

\usepackage[amsmath,thmmarks]{ntheorem}
\theoremseparator{.}
\theorembodyfont{\normalfont} 
\newtheorem{assumption}{Assumption}

\begin{document}

   \title{Ruprecht~147 DANCe\thanks{Tables \ref{table:gaia_data_and_probabilities} to \ref{table:extended_mass} are available in electronic form
at the CDS via anonymous ftp to cdsarc.u-strasbg.fr (130.79.128.5)
or via http://cdsweb.u-strasbg.fr/cgi-bin/qcat?J/A+A/}}

   \subtitle{I. Members, empirical isochrone, luminosity and mass distributions}

    \author{J. Olivares\inst{1}
          \and H. Bouy \inst{1}
          \and L.~M. Sarro\inst{6}
          \and N. Miret-Roig\inst{1}
          \and A. Berihuete\inst{9}
		  \and E. Bertin\inst{7}          
          \and D. Barrado\inst{8}
          \and N. Hu\'elamo\inst{8}
          \and M. Tamura\inst{2,3,4}
          \and L. Allen\inst{5}
		  \and Y. Beletsky\inst{10}  
          \and S. Serre\inst{1}
          \and J.-C. Cuillandre\inst{11}
}

   \institute{Laboratoire d'astrophysique de Bordeaux, Univ. Bordeaux, CNRS, B18N, allée Geoffroy Saint-Hilaire, 33615 Pessac, France.
         \email{javier.olivares-romero@u-bordeaux.fr}
         \and
         Department of Astronomy, The University of Tokyo, 7-3-1, Hongo, Bunkyo-ku, Tokyo, 113-0033, Japan
         \and
         National Astronomical Observatory of Japan, 2-21-1, Osawa, Mitaka, Tokyo, 181-8588, Japan
         \and
         Astrobiology Center, 2-21-1, Osawa, Mitaka, Tokyo, 181-8588, Japan
         \and
         National Optical Astronomy Observatory, 950 North Cherry Avenue, Tucson, AZ 85719, USA
         \and
         Depto. de Inteligencia Artificial , UNED, Juan del Rosal, 16, 28040 Madrid, Spain 
         \and
         Institut d'Astrophysique de Paris, CNRS UMR 7095 and UPMC, 98bis bd Arago, F-75014 Paris, France 
         \and
         Centro de Astrobiolog\'\i a, Dpto de Astrof\'\i sica, INTA-CSIC, ESAC Campus, Camino Bajo del Castillo s/n, E-28692, Villanueva de la Ca\~nada, Madrid, Spain
         \and
         Dept. Statistics and Operations Research, University of C\'adiz, Campus Universitario R\'\i o San Pedro s/n, 11510 Puerto Real, C\'adiz, Spain
         \and
         Las Campanas Observatory, Carnegie Institution of Washington, Colina el Pino, 601 Casilla, La Serena, Chile
         \and 
         AIM Paris Saclay, CNRS/INSU, CEA/Irfu, Universit\'e Paris Diderot, Orme des Merisiers, France 
            }

   \date{Received ; accepted }

 
  \abstract
   {Ruprecht 147 is the oldest (2.5 Gyr) open cluster in the solar vicinity (< 300 pc), making it an important target for stellar evolution studies and exoplanet searches.}
   {Derive a census of members and the luminosity, mass, and spatial distributions of the cluster.}
   {We use an astro-photometric data set including all available information from the literature together with our own observations. We process the data with an updated version of an existent membership selection methodology.}
   {{We identify 259 high-probability candidate members, including 58 previously unreported. All these candidates cover } the luminosity interval between \texttt{G}$\gtrsim$6~mag to \texttt{i}$\lesssim$21~mag. The cluster luminosity and mass distributions are derived with an unprecedented level of details allowing us to recognize, among other features, the Wielen dip.  The mass distribution in the low-mass regime drops sharply at 0.4~M$_{\sun}$ even though our data are sensitive to stellar masses down to 0.1~M$_{\sun}$, suggesting that most very-low-mass members left the cluster as the result of its dynamical evolution. In addition, the cluster is highly elongated (ellipticity $\sim$ 0.5) towards the galactic plane, and mass segregated.}
   {Our combined \textit{Gaia}+DANCe data set allows us to obtain an extended list of cluster candidate members, and to derive luminosity, mass and projected spatial distributions in the oldest open cluster of the solar vicinity.}


   \maketitle
%

\section{Introduction}
\label{section:introduction}
Discovered by John Herschel in 1833 \citep{1833RSPT..123..359H}, \object{Ruprecht 147}, also known as \object{NGC6774}, is one of the oldest open clusters in the solar vicinity \citep[2.5~Gyr at 300~pc,][]{2013AJ....145..134C}. These properties make it an important target for studies of stellar evolution and exoplanet searches. 

Its scientific potential has been  plainly revealed only recently in \citet{2013AJ....145..134C}. In their extensive astrometric, photometric and spectroscopic analysis of the cluster properties, these authors identified 108 candidate members, derived an improved distance modulus of $m-M=7.35\pm0.1$~mag, an extinction of $A_{V}=0.25\pm0.05$~mag, a super-Solar metallicity of $[M/H]=+0.07\pm0.03$ dex, and an age of $t=2.5\pm0.25$~Gyr. 

More recently \citet{Babusiaux2018} used \textit{Gaia} Data Release 2 \cite[][hereafter DR2]{2018A&A...616A...1G} to identify 154 members (after applying photometric filters to their 234 astrometric candidates) and derived a distance modulus of $7.455$, an age of $\sim$2 Gyr ($\log t = 9.3^{+0.08}_{-0.06}$) and a reddening of $E(B-V)=0.08\pm 0.04$. Shortly after, \citet{Cantat-Gaudin2018} used \textit{Gaia} DR2 to identify  191 candidate members with a mean parallax of $3.26 \pm 0.09$ ($\sim$ 307 pc). Later on, \citet{2018ApJ...866...67T} determined and age of $2.65\pm0.25$ Gyr (with an extra $\pm0.13$ Gyr uncertainty due to metallicity), and an extinction of $0.347\pm0.09$ mag for the Eclipsing Binary (hereafter EB) and cluster member \citep{2013AJ....145..134C} \object{EPIC 219394517}. In addition, \citet{2018AJ....155..173C} determined the metallicity ($[Fe/H]=+0.14\pm0.04$ dex) of the cluster member and solar twin EPIC 21980081. The previous metallicity values \citep{2013AJ....145..134C,2018AJ....155..173C} are in agreement with the latest determination of the cluster metallicity ( $[Fe/H]=+0.08\pm0.07$ dex) done by \citet{2018A&A...619A.176B} based on 21 high probability cluster members.

Furthermore, in recent years it has been found that Ruprecht 147 hosts a variety of interesting objects, as eclipsing binaries \citep{2018ApJ...866...67T,2016PhDT.......246C}, a transiting exoplanet \citep{2018AJ....155..173C}, and a transiting brown dwarf \citep{2017AJ....153..131N, 2016PhDT.......246C}.

Although the general properties of Ruprecht 147 (distance, age, metallicity and reddening) have been thoroughly determined, a detailed analysis of its mass distribution and kinematical stage is still missing. Given its proximity and advanced age, it provides a unique opportunity to benchmark our current understanding of open cluster evolution. Comparing its mass, spatial and kinematic distributions with those of younger and intermediate age clusters can shed light on the processes that shape the clusters life and ultimate dissolution in the galaxy.

The mean proper motion of the cluster reported by \citet{2013AJ....145..134C} is relatively large ($\mu_\alpha cos(\delta), \mu_\delta$)=(-1.1,-27.3) mas yr$^{-1}$, making it a  propitious target for the COSMIC-DANCe survey \citep[see for example][]{2013A&A...554A.101B,2015A&A...577A.148B}. The availability of a significant  number of archival images obtained as early as 2008 incited us to include the cluster in our target list and observations.

In this work we use \textit{Gaia} DR2 data  at the bright end, complemented with the DANCe one at the faint end to identify Ruprecht 147 members, derive empirical isochrones in standard filters, and obtain its luminosity and mass distributions\footnote{The mass distribution is the probability density function of the mass of the cluster stars. It must not be confused with the typical mass function which is given in units of stars per mass per parsec cubic.}.

In Section \ref{section:assumptions} we lay down the assumptions made to achieve our objectives. Section \ref{section:datasets} presents our data sets, their completeness limits, and the current list of members from the literature. Section \ref{section:membership_selection} introduces the improvements that we add to \citet{2014A&A...563A..45S} methodology in order to identify cluster members. In Section \ref{section:results} we present the results obtained using this improved membership selection method. Finally, in Section \ref{section:summary} we summarize our results and present our conclusions.

\section{Assumptions}
\label{section:assumptions}

This section intends to explicitly state the assumptions and simplifications made to accomplish our objectives. They are sorted in order of appearance.

\subsection{In the data set}
\label{subsection:assumptions_data}

\begin{assumption}
\label{ass:6deg}
We limit the study to a circular area of 6$^\circ$ radius around the cluster centre set at R.A.=289.10$^\circ$, Dec.=-16.38$^\circ$. The latter correspond to the mean positions of the candidate members of \citet{Babusiaux2018}. The 6$^\circ$ limit ($\sim$ 33pc) is chosen to provide a sample size that can be computationally tractable. We note that the corresponding area is four times larger than that covered by the previous studies of \citet{Babusiaux2018} and \citet{Cantat-Gaudin2018}, who analyzed the central 3$^\circ$ ($\sim$15pc) radius area.
\end{assumption}

\begin{assumption}
\label{ass:gaussian_uncertainties}
Measurements (e.g. photometric magnitudes, parallaxes, proper motions) are normally distributed, with mean at the observed value, and standard deviation as the uncertainty. The entries in the data set provide all necessary information (i.e. mean, variance and correlations) to reconstruct the distribution of measurements of the given observable.
\end{assumption}

\begin{assumption}
\label{ass:MCAR}
The missing-value mechanism, the cause behind the unobserved values in partially observed sources, is completely at random. {This assumption is not entirely correct (e.g. the missing values can be more frequent at the bright and faint limits of sensitivity). However, the correct treatment of the missing value mechanism requires knowledge of the selection function, which is currently unavailable for any of our data sets.} 
\end{assumption}

\subsection{In the membership selection}
\label{subsection:membership_selection}

\begin{assumption}
\label{ass:common_origin}
The cluster members share a common origin, thus have similar properties (e.g. age, metallicity, distance, space velocity) but with an intrinsic dispersion {or spread}.
\end{assumption}

\begin{assumption}
\label{ass:hetero_field}
The field population (i.e. sources projected in the same sky region of the cluster members) has an heterogeneous origin. Therefore, it shows a larger dispersion in its properties than that of the cluster members.
\end{assumption}

\begin{assumption}
\label{ass:disentanglment}
The cluster members can be probabilistically disentangled from the field population by means of statistical models. The quality of the classification depends on the information provided by the features used to construct the models, {and the degree of overlap between the distributions of field and clusters sources.}
\end{assumption}

\begin{assumption}
\label{ass:deconvolution}
{The observations of a given source are realizations of a (Gaussian) probability distribution centered at the (unknown) true values and with a dispersion given by the measurement uncertainties.} This assumption implies that the true distribution can be recovered by deconvolving the {uncertainty} distribution.
\end{assumption}

\begin{assumption}
\label{ass:independece_data}
The observed values and uncertainties of a source are independent of the rest of the sources. \citet{2018A&A...616A...2L} show that the astrometric uncertainties of the \textit{Gaia} DR2 are correlated, and this correlation can be expressed as function of the angular separation (see their Fig. 14 and 15), which implies that this assumption is incorrect. Nevertheless, given the large coverage of our data sets (see Assumption \ref{ass:6deg}), we consider the error introduced by this assumption to be negligible (0.0008 $\rm{[mas \cdot yr^{-1}]^2}$ in the variance of proper motions and 200 $\mu \rm{as}^2$ in the variance of parallax).
\end{assumption}

\begin{assumption}
\label{ass:GMM}
The cluster true properties are either normally distributed or can be reconstructed by a {finite} linear combination of normal distributions {(i.e. a Gaussian Mixture Model, hereafter GMM)}. This strong assumption {is an approximation that nevertheless simplifies enormously the processes of: a) deconvolving the uncertainties, and b) marginalizing over the missing-values.}
\end{assumption}

\begin{assumption}
\label{ass:independence_models}
The probability distributions of the cluster properties can be split into two independent models: the astrometric one, constructed from proper motions and parallaxes, and the photometric one constructed from luminosities in various bands and colours.
\end{assumption}

\begin{assumption}
\label{ass:populations}
{The cluster distribution is the result of a limited set of components: single stars, binaries and white dwarfs (hereafter WDs). Other types of members (e.g. AGB stars, Blue Stragglers, multiple systems) deviating from the main bulk of the distributions resulting from the aforementioned three components are assumed to result from probability distributions with insufficient probability densities to be detected by our methodology.}
\end{assumption}

\subsection{In the results}

\begin{assumption}
\label{ass:WD}
{Cluster members with magnitude \texttt{BP}$>$17~mag and colour index \texttt{G-RP}$<$0.3~mag are WDs. These values were set heuristically using the observed set of WDs in the candidate members of \citet{Babusiaux2018}}. These thresholds correspond to \texttt{r} > 18 mag, and \texttt{r-Y} < 1 mag. 
\end{assumption} 

\subsection{In the empirical isochrones}

\begin{assumption}
\label{ass:cluster_age}
The cluster age is 2.5~Gyr \citep{2013AJ....145..134C}.
We choose the value provided by \citet{2013AJ....145..134C} due to their detailed spectro-photometric analysis. This value is compatible, within the uncertainties, with those reported by \citet{Babusiaux2018} and \citet{2018ApJ...866...67T}.
\end{assumption}

\begin{assumption}
\label{ass:cluster_metallicity}
The cluster metallicity is $Z/Z_{\odot}=0.017$ \citep{2018A&A...619A.176B}. As mentioned before, this value is in agreement with the previous determinations of \citet{2013AJ....145..134C,2018AJ....155..173C}.
\end{assumption}

\subsection{In the luminosity and mass distributions}
\begin{assumption}
\label{ass:isochrones}
Theoretical isochrones provide the true relation between the photometric magnitudes and the star mass and luminosity.
\end{assumption} 

\begin{assumption}
\label{ass:IFMR}
The mass of the WD progenitor can be recovered from the WD mass and the WD Initial-to-Final mass relation of \citet{2018ApJ...866...21C}.
\end{assumption} 

\subsection{In the projected spatial distribution}
\begin{assumption}
The uncertainties in the star position (equatorial coordinates) are negligible.
\end{assumption}

\section{Data sets}
\label{section:datasets}

\subsection{Gaia DR2 data set}
\label{subsection:gaia_dataset}

We downloaded\footnote{http://tapvizier.u-strasbg.fr/adql/}  \textit{Gaia} DR2 astrometric and photometric measurements for all sources contained within a 6$^\circ$ radius around the cluster center {(see Assumption \ref{ass:6deg})}. Approximately 4 million sources (out of 16 million) only have positions and photometric measurements. Although a classification can still be done based only on photometric measurements, this will be highly contaminated by the field population that interlopes the cluster sequence in the colour-magnitude diagrams. Therefore, we decided to discard these sources. Our \textit{Gaia} DR2 data set, in the following referred as GDR2, contains proper motions, parallax and  photometry (\texttt{BP,G,RP}) of $\sim$ 12 million objects.

\subsection{DANCe data}
\label{subsection:dance_dataset}
DANCe (Dynamical Analysis of Nearby Clusters) is a multi-instrument survey that combines public archival data (i.e. catalogues and images) with our own observations, and derives comprehensive astrometric and photometric (visible and near-infrared) catalogues of sources in nearby (<500pc) clusters \citep[see for example][]{2013A&A...554A.101B,2015A&A...575A.120B,2015A&A...577A.148B}.

In this article, we present our astrometric and photometric analysis of a  8\degr$\times$ 6\degr region centered on Ruprecht 147, which is based on the combination of archival and new wide field images obtained between 2008 and 2017. 

We searched:
\begin{itemize}
\item the European Southern Observatory (ESO) archive
\item the National Optical Astronomy Observatory (NOAO) archive
\item the PTF archive hosted at the NASA/IPAC Infrared Science Archive (IRSA)
\item the Canadian Astronomy Data Centre (CADC) archive
\item the Isaac Newton Group (ING) archive 
\item the WFCAM Science (WSA) archive
\end{itemize} 
for wide field images within a 8\degr$\times$ 6\degr region centered on Ruprecht 147. The data found in these public archives were complemented with our own observations using the Las Campanas Swope telescope and its Direct CCD camera \citep{Swope_NewCCD} and  the  Dark Energy Camera \citep[DECam,][]{DECAM} mounted on the Blanco telescope at the Cerro Tololo  Inter-American Observatory. Table~\ref{tab:observations} gives an overview of the various cameras used for this study. The \texttt{u}-band images obtained with VST/OmegaCam and CTIO/DECam available in the archives were not included in  the astrometric analysis because of their high sensitivity to atmospheric refraction, but the corresponding photometry is included in the final catalogue.  

\begin{figure}[htp]
\begin{center}
\includegraphics[width = 0.5\textwidth]{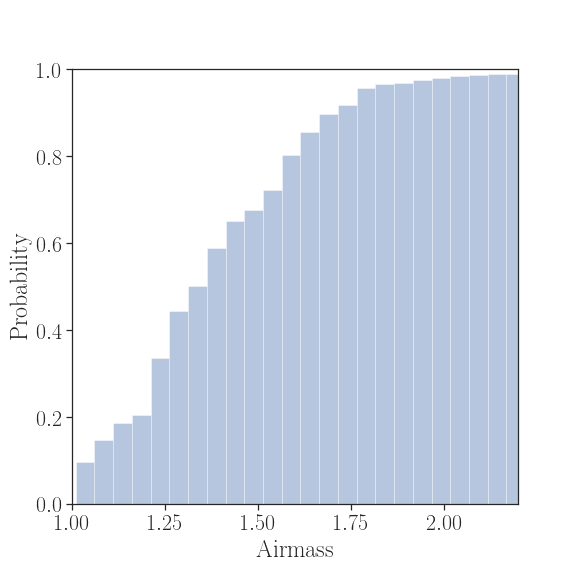}
\includegraphics[width = 0.5\textwidth]{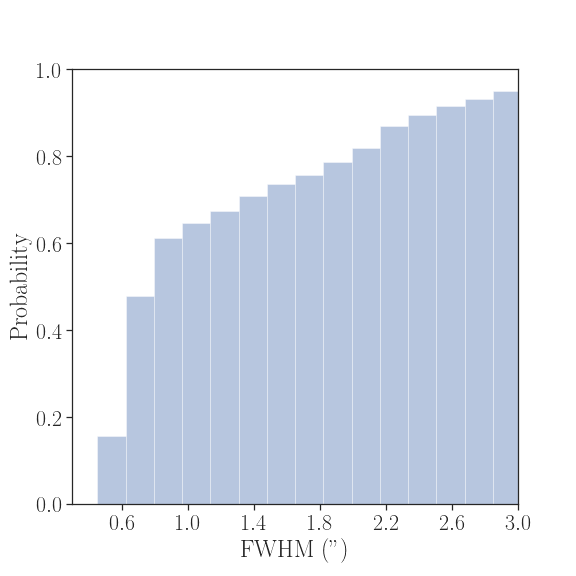}
\caption{Top: Cumulative distribution of airmass for the observations. Bottom: Cumulative distribution of average FWHM for the images. }
\label{fig:airmass_fwhm}
\end{center}
\end{figure}

The top panel of Figure~\ref{fig:airmass_fwhm} shows the cumulative distribution of images as a function of airmass for the observations used in this study. About 70\% of the observations were obtained at airmass $\le$1.5. Ruprecht~147 is located at Dec. $\sim$ -16\degr, explaining the rather flat distribution of airmass as we gathered data obtained from both hemispheres. The bottom panel of Figure~\ref{fig:airmass_fwhm} shows the cumulative distribution of images as a function of the average full-width at half maximum (FWHM) measured for all individual unresolved detections (point sources) in the images. About 62\% have FWHM $\le$1\arcsec. 

In all cases except for CFHT/MegaCam, DECam and UKIRT images, the raw  data and associated calibration frames were downloaded and processed using standard procedures  using an updated version of \emph{Alambic} \citep{2002SPIE.4847..123V}, a software suite developed and optimized for the processing of large multi-CCD imagers. In the case of CFHT/MegaCam, the  images processed and calibrated with the \emph{Elixir} pipeline were retrieved from the CADC archive \citep{Elixir}. In the case of DECam, the images processed with the community pipeline \citep{2014ASPC..485..379V} were retrieved from the NOAO public archive. UKIRT images processed by the Cambridge Astronomical Survey Unit were retrieved from the WFCAM Science Archive.

\begin{table*}
\caption{Instruments used in this study \label{tab:observations}}

\begin{tabular}{lccccccclc}\hline\hline
Observatory   & Instrument        & Filters              & Platescale   & Field of view & Epoch Min./Max.  & Ref. \\
                      &                          &                         & [\arcsec pixel$^{-1}$] &                      &                  &                     &                           &      \\
\hline
CTIO (Blanco)  & DECam   & $g,r,i,Y$  & 0.26 & 1.1\degr radius & 2012--2017  & (1) \\
CFHT   & MegaCam   & $g,r,i,z$  & 0.18 & 1\degr$\times$1\degr & 2008  & (2) \\
INT & WFC & $u,v,b,\beta,y$ (Str\"omgren) & 0.33 & 34\arcmin$\times$34\arcmin & 2015 & (3) \\
UKIRT &  WFCAM & $J,Ks$ & 0.4 & 40\arcmin$\times$40\arcmin \tablefootmark{a} & 2011 & (4)\\
LCO Swope & Direct CCD & $i$ & 0.43 & 15\arcmin$\times$14\arcmin & 2013 & (5) \\
VST & OmegaCam & $r,i,z$ & 0.21 & 1\degr$\times$1\degr & 2013--2016 & (6) \\
Palomar 48" & PTF & $g,r$ & 1.0 & 3\fdg3$\times$2\fdg2\tablefootmark{b} & 2010--2012 & (7) \\
VISTA & VIRCAM & $J,Ks$ & 0.34 & 1\fdg3$\times$1\fdg0 \tablefootmark{a}   & 2013--2014 & (8) \\
\hline
\end{tabular}
\tablefoottext{a}{Large gaps between detectors, partial coverage of the focal plane}
\tablefoottext{b}{One of the 12 detectors is dead}

\tablebib{(1) \citet{DECAM}, (2) \citet{CFHT_MegaCam}, (3) \citet{INT_WFC}, (4) \citet{UKIRT_WFCAM}, (5) \citet{Swope_NewCCD}, (6) \citet{VST_OmegaCam}, (7) \citet{PTF}, (8) \citet{VISTA}}

\end{table*}

\subsubsection{Astrometric analysis}
After a visual rejection of problematic images (mostly due to loss of guiding, tracking or read-out problems), the dataset included 2\,068 individual images originating from 8 cameras. A total of 303\,195\,115 individual sources were detected in these images and the final catalogue includes  10\,201\,564 unique sources.

The astrometric calibration was performed as described in \citet{2013A&A...554A.101B}, with the following updates:
\begin{itemize}
\item Support for differential geometry maps to correct for the "tree rings" and the edge distortions in thick, back-illuminated, fully-depleted CCD  \citep[e.g. DECam,][]{2006SPIE.6276E..08D}
\item improved cross-matching algorithm with HEALPix tessellation \citep{2005ApJ...622..759G}
\item improved field grouping (now based on exact footprints)
\item added support for \textit{Gaia} DR2 \citep{gaiadr2} as an astrometric reference
\item increased astrometric reference query limit to 100 million sources per group
\end{itemize} 

The recently released \textit{Gaia} DR2 catalogue was therefore used as external astrometric reference instead of the 2MASS catalogue, leading to a much improved astrometric solution. The final average internal and external 1-$\sigma$ residuals were estimated to be $\sim$15~mas for high signal-to-noise (photon noise limited) sources. 

As explained in \citet{2013A&A...554A.101B}, the proper motions computed are relative {to one another} and display an offset with respect to the Geocentric Celestial Reference System. {We estimate the offsets by computing the median of the difference between our proper motions and those of the \textit{Gaia} DR2 } after rejecting outliers using the modified Z-score \citep{iglewicz}. We find offsets of $(\Delta\mu_{\alpha}\cos(\delta),\Delta\mu_{\delta})=(3.72,5.63)$~mas/yr. The uncertainty on these offsets, estimated using bootstrapping, is $<$0.004~mas/yr.

\subsubsection{Photometric analysis}
The photometric calibration was performed only for the \texttt{g,r,i,z,y} and \texttt{J,H,Ks} images, and was not attempted for the INT Str\"omgren images. In the case of PTF images, the \texttt{g,r} photometry was not used in the photometric analysis because of their significantly coarser pixel scale. 

\subsubsection*{Individual images}
The photometric zero-point of all individual images obtained in the \texttt{u,g,r,i,z,y, H$\alpha$} and \texttt{J,H,Ks} filters was computed by direct comparison of the instrumental {\sc SExtractor} \verb|MAG_AUTO| magnitudes with an external catalogue: 
\begin{itemize}
\item \texttt{J,H,Ks} images were tied to the 2MASS catalogue \citep{2MASS}
\item \texttt{u, H$\alpha$} images were  tied to the VPHAS+ photometry \citep{VPHAS}
\item \texttt{g,r,i,z,y} images were tied to the Pan-STARRS PS1 release \citep{ps1}
\end{itemize}
The individual zero-points were computed as follows: first, clean point-like sources were selected based on the following criteria:
\begin{itemize}
\item {\sc SExtractor}  $|$\verb|SPREAD_MODEL|$|\le$0.02. The \verb|SPREAD_MODEL| is a morphometric linear discriminant parameter obtained when fitting a Sérsic model, which is useful to classify point-sources from extended sources \citep[see e.g.][]{2013A&A...554A.101B}.
\item a difference between the Kron and PSF instrumental magnitudes smaller than 0.02~mag ($|$\verb|MAG_AUTO|-\verb|MAG_PSF|$|\le$0.02~mag), used as an additional test to reject extended or problematic sources
\item {\sc SExtractor} \verb|FLAG|=0, to avoid any problem related to blending, saturation, truncation
\end{itemize}
The closest match within 1\arcsec\, to these clean point-like sources was then found in the reference catalogue. The zero-point was computed as the median of the difference between the reference and instrumental (\verb|MAG_AUTO|) magnitudes. The median absolute deviation are typically of the order of 0.01$\sim$0.09~mag depending on the filter.

\subsubsection*{Deep stacks}
We median-combined all the images obtained with the same camera and in the same filter to obtain a deep stack and extract the corresponding photometry. These stacks made of many epochs were not used for the astrometric analysis but allowed us to significantly improve the detection sensitivity in all filters, and recover or improve the photometry of faint sources obtained in the individual images. 

\subsubsection{External astrometry and photometry}
\label{subsection:DANCe_final_catalogue}
As in \citet{2013A&A...554A.101B}, the photometry and astrometry extracted from the images is complemented by that reported in external catalogues. The recent \textit{Gaia} DR2 catalogue provides accurate parallaxes and proper motions for sources up to $G\sim$20.7~mag. We retrieved all the sources reported in the \textit{Gaia} DR2 catalogue over the area covered by our images:
\begin{itemize}
\item[] 285.08\degr$\le$R.A.$\le$293.0\degr
\item[] -19.55\degr$\le$Dec.$\le$-13.48\degr
\end{itemize}
and cross-matched with our catalogue using a 1\arcsec\, radius. We find that our catalogue includes 3\,960\,090 sources not reported in \textit{Gaia} DR2, mostly because they are beyond the \textit{Gaia} sensitivity limit. However, 169\,063 of these sources have 14$<$\texttt{r}$<$20~mag, well within both the \textit{Gaia} and DANCe sensitivity limits. According to the previous value, we estimate that \textit{Gaia} is missing 3\% of the sources. The latter represents a completeness of 97\%, which is higher than the typical value of $\sim$90\% reported by \citet[][see their Fig. 7]{2018A&A...616A..17A} for a density of 10$^5$ (as that of our data set). On the other hand, a total of 2\,820\,764 sources from the \textit{Gaia} DR2 catalogue have no counterpart in our catalogue, either because they are saturated in our images or because they fall in areas of the region defined above not covered by any image in our data set. 

Because \textit{Gaia}'s proper motion measurements are much more accurate and reliable than our ground-based measurements, we always prefer them over our own measurements, and keep our proper motion measurements only for sources without counterparts in the \textit{Gaia} DR2 catalogue. 

\begin{figure}[htp]
\begin{center}
\includegraphics[width = \columnwidth]{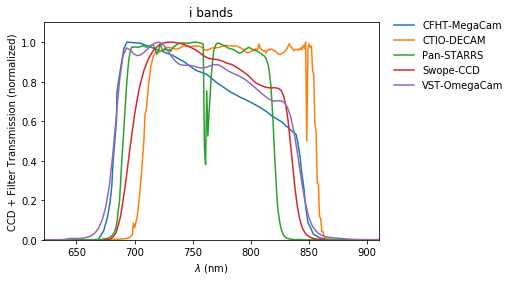}
\caption{Transmission curves of the $i$ filters used in this study. Similar differences exist for the other bands.}
\label{fig:iband}
\end{center}
\end{figure}

Given the heterogeneous spatial coverage of the various datasets, we complement the combined DANCe+\textit{Gaia} DR2 catalogue obtained above (and including \texttt{u,g,r,i,z,y,J,H,Ks,G,BP,RP}) with Pan-STARRS (\texttt{g,r,i,z,y}), 2MASS (\texttt{J,H,Ks}) and ALLWISE \citep[all 4 bands,][]{WISE} photometry. The corresponding photometric measurements were either added to our catalogue when no measurement was available for the corresponding source in our data, or the weighted average of all (our and external catalogues) measurements was computed. In some cases, significant differences exist among the filters of the same photometric band. For example, Fig. \ref{fig:iband} shows the transmission curves of all \texttt{i} band filters. These systematic differences contribute to the noise of the photometric dispersion in the colour-magnitude and colour-colour diagrams used for our analysis. Moreover recent variability studies of young clusters found typical amplitudes of 0.03~mag \citep[e.g.][]{2016AJ....152..113R,2018AJ....155..196R}. Since our aim is to identify cluster members, and to take into account both the systematic differences among filters of the same band and the intrinsic photometric variability, we conservatively add, in quadrature, 0.05~mag to all the photometric uncertainties in our catalogue.

Figure \ref{fig:completeness} shows the distributions of \texttt{grizyJHKs} magnitudes in our final catalogue. Given the low extinction in the area, the maximum value of the magnitudes gives an estimate of the limit of sensitivity of the survey. It is nevertheless important to remember that the spatial coverage of the various instruments is not homogeneous and the {depth of the survey varies spatially}. The limit of sensitivity of the external catalogues merged with our data (e.g 2MASS and Pan-STARRS) are sometimes visible as secondary maxima. 


\subsection{Completeness of the data sets}
\label{subsection:completeness}

The completeness of our GDR2 data set is inherited from the \textit{Gaia} DR2 catalogue. Since we only removed those sources without proper motions, the completeness corresponds to that of the 5-parameter solution of the \textit{Gaia} DR2. As reported in Table B.1 of \citet{2018A&A...616A...2L} the completeness of this 5-parameter solution is more limited than that of the entire catalogue, reaching 94.5\% at 19~mag in \texttt{G}, and dropping to 82.9\% at 20 \texttt{G} band. We therefore assume our analysis to be complete up to \texttt{G}=19~mag, which roughly corresponds to 19.5~mag in \texttt{BP} and 18.4~mag in \texttt{RP}. At the bright side, the \textit{Gaia} data is complete down to \texttt{G}= 7~mag \cite[see Sect. 6.2 of][]{2018A&A...616A...1G}.

Given its heterogeneous origin, the astrometric and photometric completeness of the DANCe data set has a complex distribution. We tentatively define a central region of the cluster where the high density and coverage of images must ensure homogeneous photometric and spatial completeness. This region, referred hereafter as the "DANCe Central" data set, includes all DANCe sources within:
\begin{itemize}
    \item[] 287.4\degr$\leq \rm{R.A.} \leq$  290.4\degr,
    \item[] -17.83\degr$\leq \rm{Dec.} \leq$ -14.98\degr.
\end{itemize}

In Table \ref{tab:completeness} we give the percentage of DANCe and DANCe Central sources observed in each photometric band. As can be seen, in both data sets the best coverage comes from the \texttt{grizY} photometry.

\begin{table}[ht]
\centering
\caption{Percentage of sources with observed photometry in the DANCe and DANCe Central data sets.} 
\label{tab:completeness}
\resizebox{\columnwidth}{!}{%
\begin{tabular}{lrrrrrrrrr}
  \hline
Data set &\texttt{u} & \texttt{g} & \texttt{r} & \texttt{i} & \texttt{z} & \texttt{Y} & \texttt{J} & \texttt{H} & \texttt{Ks} \\ 
  \hline
DANCe         & 8\% & 73\% & 83\% & 94\% & 85\% & 82\% & 48\% & 16\% & 36\% \\ 
DANCe Central & 14\% & 59\% & 75\% & 95\% & 75\% & 67\% & 45\% & 8\% & 32\% \\ 
   \hline
\end{tabular}

}
\end{table}

\begin{figure*}
\begin{center}
\includegraphics[width = \textwidth]{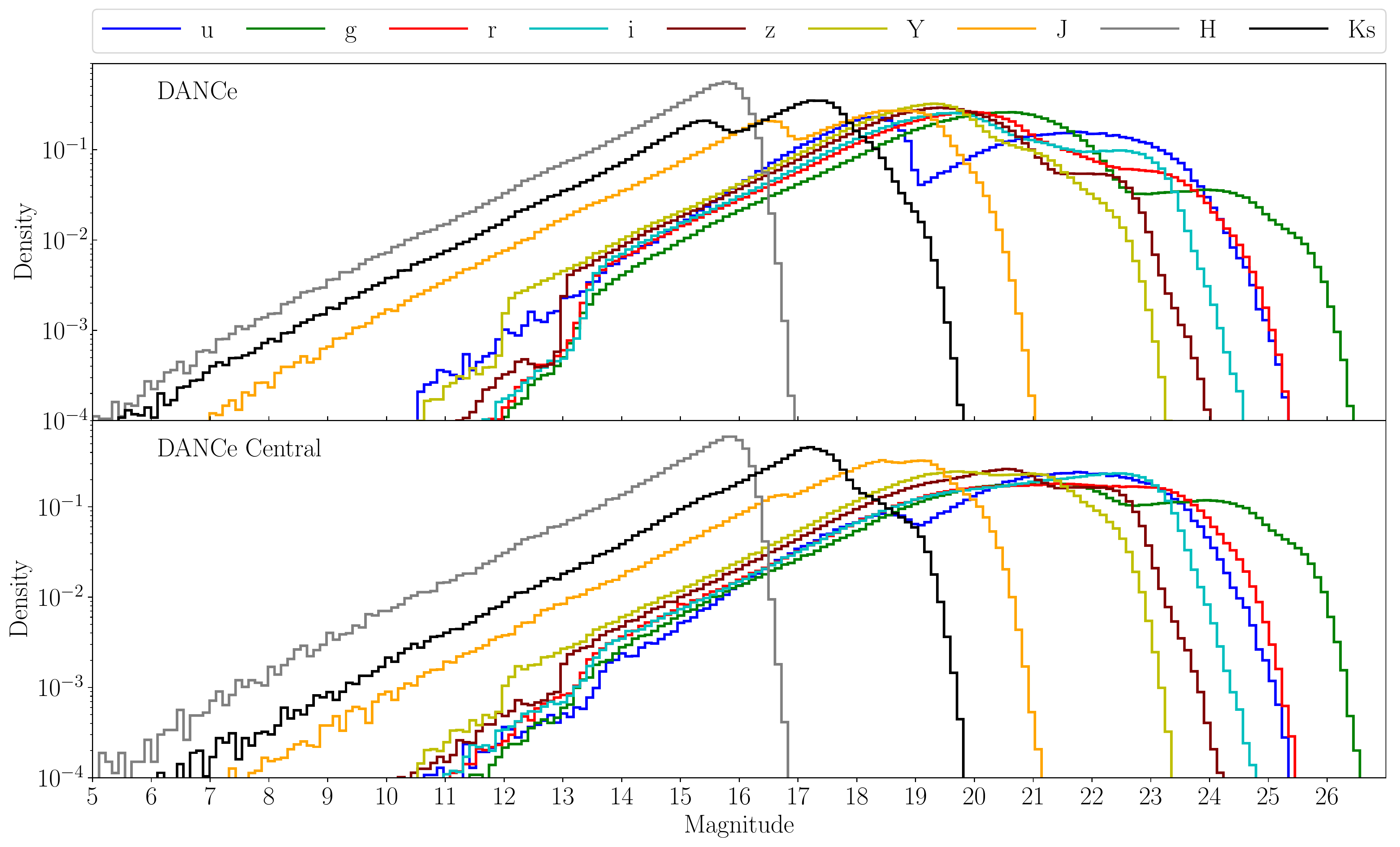}
\caption{Density of sources as function of the magnitude for the DANCe (top panel) and DANCe Central (bottom panel) data sets.}
\label{fig:completeness}
\end{center}
\end{figure*}

For both the DANCe and DANCe Central data sets, we define the completeness limit of each photometric band (independently of the others) as the magnitude at which the density of sources reaches a maximum. Figure~\ref{fig:completeness} shows the density of sources in the DANCe and DANCe Central data sets as function of magnitude. {The \texttt{ugriz} photometry is severely limited below $\sim 13$~mag, where our images saturate. At the faint end, multiple local maxima corresponding to the sensitivity limits of the diverse instruments can often be seen.}


From Fig. \ref{fig:completeness}, we estimate the completeness intervals that are reported in Table \ref{table:completeness}.

\begin{table}[ht!]
    \centering
    \caption{Interval of completeness for the DANCe and DANCe Central data sets.}
    \label{table:completeness}
    \begin{tabular}{c|cc|cc}
    Band & \multicolumn{2}{c}{DANCe} & \multicolumn{2}{c}{DANCe Central} \\
         & lower~mag& upper~mag& lower~mag& upper~mag\\
    \hline
    \texttt{u}    & 13.0   & 18.3 & 13.0 & 21.7 \\
    \texttt{g}    & 13.5   & 20.6 & 13.5 & 21.4 \\
    \texttt{r}    & 13.5   & 20.0 & 13.5 & 21.4 \\
    \texttt{i}    & 13.5   & 19.7 & 13.5 & 22.4 \\
    \texttt{z}    & 13.0   & 19.4 & 13.0 & 20.5 \\
    \texttt{Y}    & 12.0   & 19.3 & 12.0 & 19.8 \\
    \texttt{J}    & 7.0    & 18.7 & 7.0  & 18.4 \\
    \texttt{H}    & 6.0    & 15.8 & 6.0  & 15.9 \\
    \texttt{Ks}   & 6.0    & 17.2 & 6.0  & 17.2 \\
    \hline
    \end{tabular}
    
\end{table}

\subsection{Previously known candidate members}
\label{subsection:previous_members}
As mentioned in Section~\ref{section:introduction}, the candidate members of Ruprecht 147 have been analyzed in mostly three studies: \citet{2013AJ....145..134C,Babusiaux2018} and \citet{Cantat-Gaudin2018}. The latter two used the \textit{Gaia} DR2 data and therefore offer a straightforward comparison to our results. We cross-identify their lists of members with our data sets. As expected the 234 astrometric candidate members of \citet{Babusiaux2018} and the 191 candidates of \citet{Cantat-Gaudin2018} are all included in our GDR2 data set. However, all but one of the 108 candidate members of \citet{2013AJ....145..134C} have a counterpart in our GDR2 data set (the remaining one, \object{2MASS J19160785-1610360} has no measurements in \textit{Gaia} DR2). When cross-identifying the candidates from the literature with our DANCe data set, we recover all  \citet{2013AJ....145..134C} and  \citet{Babusiaux2018} candidate members, but only 190 from the 191 from  \citet{Cantat-Gaudin2018}, the remaining one is at the border of our DANCe region.

\section{Membership selection}
\label{section:membership_selection}

In the literature there are several methodologies that deal with the detection of cluster members \cite[e.g.][]{2018A&A...617A..15O,2018ApJ...856...23G,2018A&A...615A..49C,2014A&A...563A..45S}. However, neither of them allow to extract candidate members from data sets of millions of sources while combining astrometric and photometric features and their uncertainties and correlations. For example, the methodology of \citet{2018A&A...615A..49C} include correlations in the uncertainties but do not use the photometric measurements. On the other hand, the methodology of \citet{2018ApJ...856...23G} takes into account the source photometry to constrain the prior of the parallax. However, it does not take into account correlations in the uncertainties. The methodology of \citet{2014A&A...563A..45S} is able to work on data sets of millions of sources taking into account both astrometric and photometric informations, but does not include the observational uncertainties in the construction of the cluster model. In our opinion, the methodology of \citet{2018A&A...617A..15O} has clear advantages over the others: takes into account the full covariance matrix of the observations, incorporates photometry, and deals with partially observed objects. However it is so computationally expensive that it is impractical for our data set made of millions of sources.
 
For all these reasons, we decided to improve the methodology  of \citet{2014A&A...563A..45S} incorporating uncertainties and their correlations to extract candidate members from data sets of millions of sources. In the following we present a brief summary of the original method of \citet{2014A&A...563A..45S}, and refer the interested reader to the mentioned work for more details. Afterwards, we introduce our improvements.
 
Not all the combination of astrometric measurements, colours and luminosities are equally informative to identify members among field sources and the first step consists in choosing an optimized representation space among all the available observables. Second, the data set is split into the completely observed sources (hereafter the completely-observed data set), and the incompletely observed ones, i.e. objects with missing measurements in some bands (hereafter the partially-observed data set). Third, the cluster and field populations are modeled using the completely-observed data set. The field model uses a GMM constructed in the entire representation space. The cluster model is the independent product of a GMM constructed in the proper motion space and a multivariate Gaussian in the photometric part of the representation space with the mean described by a principal curve (representing the cluster sequence). The field GMM remains fixed throughout the inference process, while the cluster proper motions GMM and principal curves are iterated until convergence. The first iteration requires an initial set of candidate members upon which the cluster model is constructed. The cluster and field models are used to assign Bayesian membership probabilities to all completely-observed sources. A new classification (i.e. into field and cluster) is performed based on a probability threshold called $p_{in}$. This probability threshold must be set a priori by the user. Since this value establishes how conservative or lax the cluster model is, it can be set according to the user objectives {(lower $p_{in}$ values result in more inclusive cluster definition sets and, possibly more contamination from field sources as well).} The new list of cluster candidate members found with the chosen $p_{in}$ is then used as the starting point of the next iteration. The methodology is said to converge when no more cluster candidates are added or removed in successive iterations. Finally, the learned cluster and field models are used to compute the membership probabilities of the missing-value sources and the final classification is performed.

\subsection{Modifications}
\label{subsection:modifications}
In the present work we undertake a number of improvements to the methodology of \citet{2014A&A...563A..45S} described above. We modify it to take into account the properties of both the \textit{Gaia} DR2 data and the cluster Ruprecht 147. These modifications include the use of parallax measurements, of all uncertainties and correlations (when available) in the construction of the cluster model, and a new model to describe the photometry of WDs.
 
We model the cluster 3D space of proper motions and parallaxes with a GMM, which is independent of the photometry and takes into account the full covariance matrix of the uncertainties (correlations included). As in  \citet{2014A&A...563A..45S}, the number of Gaussian components is set using the Bayesian Information Criterion \citet[][hereafter BIC]{Schwarz1978}. In the field model, we include the parallax as any other feature (i.e. allowing for correlations in the multidimensional GMM).

The uncertainties and their correlations are incorporated by means of full covariance matrices and deconvolved from the the cluster intrinsic dispersion using the \emph{ExtremeDeconvolution} package \citep{2011AnApS...5.1657B}. The latter uses the Expectation-Maximization algorithm to obtain the parameters of a GMM in which the individual uncertainties (in the form of a full covariance matrix if available) are deconvolved to obtain the intrinsic covariance matrices of the GMM.
In a similar way, the photometric uncertainties are deconvolved from the cluster intrinsic photometric dispersion. In both photometric and astrometric models, the value of the \texttt{w} parameter\footnote{{The \texttt{w} parameter sets the minimum variance allowed in the covariances of the GMM, thus avoiding the proposal of singular covariance matrices.}}, required by the \emph{ExtremeDeconvolution} package, is set to the square of the minimum uncertainty present in the data set. 

To model the cluster population photometry we modify the \citet{2014A&A...563A..45S} methodology as follows. WDs are included in the cluster model by means of a Gaussian in the photometric space. The parameters of this Gaussian are inferred from the WDs found in the list of candidate members at each iteration (WDs are identified using Assumption \ref{ass:WD}). In addition, the fractions of binaries and WDs do not remain fixed but are updated at each iteration. Binaries are selected as those candidate members whose magnitudes are 0.2~mag brighter than the cluster photometric sequence (i.e. the principal curve at that iteration). In \citet{2014A&A...563A..45S} the density of members along the principal curve was modelled by a simple non-parametric kernel density estimate (herafter KDE) over the $\lambda$ parameter, which represents the length along the curve from an arbitrary but fixed origin. We modified this density of members along the principal curve and represent it as a mixture density comprising the original KDE plus a uniform distribution. The latter ensures that the model can detect new members in poorly populated regions of the cluster sequence, where the simple KDE yields low membership probabilities. In doing so, it avoids that the iterative procedure, which initiates with the brighter sources from the literature, stops before the empirical photometric sequence reaches the fainter end of the cluster. Once these low-density-region members are located, we remove the uniform density from the mixture modelling the members density, and run again the methodology with these members as input. This modification was not needed in \citet{2014A&A...563A..45S} since the Pleiades cluster is {an order of magnitude richer than Ruprecht 147}, and the previously known members covered most of the cluster sequence.

In the methodology of \citet{2014A&A...563A..45S} the authors use a $p_{in}$ value in the range between 0.95-0.99, which is highly conservative.
While restrictive (high) $p_{in}$ values may hinder the extension of the cluster sequence to the faint end (despite the modification described previously), lower ones may introduce field contaminants in the cluster definition set, and thus prevent convergence of the algorithm. 
Because of the small size of Ruprecht 147 ($\sim 200$ members), we decide to use less restrictive value of $p_{in}$ in the range 0.5-0.9. We start from the lower theshold ($p_{in}=0.5$)
and if after 30 iterations the code does not
converge, we increase the $p_{in}$ value until
convergence is attained. Convergence is now defined when two consecutive iterations have a relative difference between the logarithm of the cluster likelihood less than a user defined tolerance.

\subsection{Steps of the membership selection}
\label{subsection:steps}
The following is a step-by-step description of our membership selection.

\subsubsection{Preprocessing}
\label{subsection:preprocessing}
We modify the data sets as follows:
\begin{itemize}
\item Objects with large photometric uncertainties (>0.5 mag) or large proper motion uncertainties (> 100 mas yr$^{-1}$) were masked as missing. This avoids contaminating the field model (constructed from the completely-observed data set) with objects affected by extremely large uncertainties (e.g. photometric uncertainties reaching up to 29~mag in the GDR2 data set).

\item As mentioned previously, the photometric uncertainties were increased by quadratically adding 0.05~mag. This avoids discarding probable cluster members with photometric variability that would otherwise be classified as field population due to to the high precision uncertainty of the \textit{Gaia} photometric measurements. It also mitigates the effect related to differences in the various instruments transmissions.
\item Missing correlation terms were replaced by zeros.
\end{itemize}

After applying the preprocessing procedure, the GDR2, DANCe and DANCe Central data sets contain 12\,081\,778, 6\,776\,857, and 2\,246\,600 sources, respectively.

\subsubsection{Initial list of candidate members}
\label{subsection:inital_list}
We use the 191 candidate members of \citet{Cantat-Gaudin2018} as our initial list of members. We choose it because it is larger than the 154 filtered candidates of \citet{Babusiaux2018}. 

We complement the candidate members of \citet{Cantat-Gaudin2018} with the 10 WDs in \citet{Babusiaux2018}, see Assumption \ref{ass:WD}. 

\subsubsection{Representation space}
\label{section:RS}
The representation space is chosen using the method described in \citet{2014A&A...563A..45S}. However, we always include proper motions and parallaxes {(if present)} in the representation space because in our experience these are discriminant features. 

To select the photometric features of our representation spaces we use the members and non-members in each of the completely-observed data sets to run random-forest classifiers \cite[R package \emph{RandomForest},][]{Breiman2001}. The latter were trained ten times with a maximum number of 3 features and 1000 trees. We select features based on both their importance (given by the mean decrease in the Gini impurity\footnote{The Gini impurity is a measure of misclassification. The intereseted reader can find more details in \citet{Breiman2001}.}) and its percentage of observed sources, preferring those with less missing values. The only condition to choose features is that they must be linearly independent, otherwise it results in redundant information.

After testing several representation spaces (see Appendix \ref{appendix:sensitivity}) we converged to the following ones. 

For GDR2 we use:
\begin{equation}
RS_0 :\texttt{pmra, pmdec, parallax, BP, BP-G, G-RP}. \nonumber
\end{equation}

For DANCe and DANCe Central we use
\begin{align}
&RS_1:\texttt{pmra, pmdec, i, g-i, g-Y, g-Ks}. \nonumber \\
&RS_2:\texttt{pmra, pmdec, r, g-z, r-Y}. \nonumber
\end{align}

{Unfortunately, the $RS_1$ does not contain the WDs due to their missing \texttt{Ks} photometry. Thus, we complement the candidate members resulting from the $RS_1$ with the WDs of $RS_2$.}

\subsubsection{Completely and partially observed data sets}

We transform the observables and uncertainties in our data sets into the features of the representation space, and split them in the partially-observed and completely-observed data sets. 

The following steps apply only to the completely-observed members and non-members, except for the final classification of sources, which is done in the entire data set. 

\subsubsection{The field model} 
The parameters of the field GMM are inferred using the \textit{scikit-learn} package \citep{scikit-learn}. We apply the latter to ten random samples of one million elements each. These are drawn from the non-members sources in the completely-observed data set. For each of these samples, we infer the GMM parameters with varying number of components. Then, we compute the mean BIC and its standard deviation (using the ten random samples) for each number of components. We found minimum mean BIC values at 125, 150, and 140 components for the GDR2 and DANCe Central and DANCe data sets, respectively. However, these mimima are compatible within the BIC standard deviation with simpler models having 80, 100 and 120 components for the GDR2 and DANCe Central and DANCe data sets, respectively. We prefer simpler models to reduce both computational burden and possible over-fitting. Nevertheless, these simpler models represent an improvement with respect to the 26-components  field model built by \citet{2014A&A...563A..45S} from a sample of only 40\,000 sources.

\subsubsection{The cluster model} 
\label{subsection:iterative_classification}

The following steps are iteratively applied until convergence.
\begin{enumerate}

\item {Cluster model construction:} 
The cluster model, as mentioned above, is made up of the astrometric and photometric models. 

The astrometric model is a GMM whose number of components is selected, within the range 1 to 4, by means of the BIC. The parameters of this GMM are inferred from the current list of members and weighted by membership probability (set to 1 for the first iteration).

The photometric model contains three sub-models corresponding to single stars, binaries and WDs. Each of them is a multivariate Gaussian function of the photometric features of the representation space. They have the following particularities:
\begin{itemize}
\item For single stars, the mean and covariance matrix are prescribed as in \citet{2014A&A...563A..45S}, with the exception that instead of computing a simple weighted covariance matrix with the 10 closest neighbours at each point in the principal curve, the new covariance matrix deconvolves the uncertainties of these 10 neighbours by means of the \emph{ExtremeDeconvolution} package. 

\item Binaries are treated as in \citet{2014A&A...563A..45S}. 

\item For WDs, the mean and covariance matrix of the multivariate Gaussian are found using the \emph{ExtremeDeconvolution} package applied to the list of WD members (i.e. those candidate members fulfilling the conditions of Assumption \ref{ass:WD}). 
\end{itemize}

The fractions of these three models are computed from the total number of members at each iteration.

\item {Membership probability assignment:} The cluster and field likelihoods are computed for each source in the completely-observed data set using the corresponding models. Then, with the previous likelihoods and the fraction of members and non-members as priors, the cluster membership probability of each source is computed as the ratio of cluster to total (cluster plus field) posterior probability \cite[see Eq. 7 of][]{2014A&A...563A..45S}. 

\item {Reclassification:} Based on a probability threshold, $p_{in}$, established a priori, members and non-members are reclassified according to their membership probability in the current iteration. Objects with $p > p_{in}$ are considered members.  
In all data sets we run our methodology with $p_{in}$ values in the range 0.5-0.9, at steps of 0.1. Appendix \ref{appendix:sensitivity} shows the sensitivity of our results to these $p_{in}$ values.
\end{enumerate}

Convergence is achieved when the relative difference in the cluster log likelihood of two consecutive iterations is smaller than $10^{-4}$. 

\subsubsection{High probability members: defining the probability threshold}
\label{subsection:mockdata_analysis}

To classify sources into the cluster and field classes we need an optimum probability threshold, which might not be the same as the $p_{in}$. To objectively set this optimum, we analyze the quality of the classifier when applied over synthetic data sets. 

We use the learned cluster and field models to generate five synthetic data sets with properties mimicking the completely-observed sample. The synthetic field data set contains the same number of sources as the real one, while for the synthetic cluster sample we use 1000 synthetic members. Although this figure is larger than the actual number of cluster members, it is still negligible compared to the millions of non-members, and it improves the statistics of the classifier quality measures.

The uncertainties of these synthetic sources were generated by sampling from the true uncertainties, with no correlations included. Then, we run the iterative classification using exactly the same parameters as those used for the real data sets. 

Upon convergence, we analyze the classifier quality computing the confusion matrix (i.e true positives, hereafter TP, false positives, hereafter FP, false negatives, hereafter FN, and true negatives hereafter TN) and the following indicators: True Positive Rate (TPR), Contamination Rate (CR) and Distance to perfect classifier (DST), which are defined as:

\begin{align}
TPR&=\frac{TP}{TP+FN}, \\
CR&=\frac{FP}{FP+TP}, \\
DST&=\sqrt{(CR-0)^2 + (TPR -1)^2}. \label{eq:DST}
\end{align}

We do not measure the quality of the classifier based on indicators that depend on the number of true negatives. Given the size of the data set and the class unbalance (millions of non-members versus only a thousand members), these indicators return flat values for almost all probability thresholds (e.g. the false positive rate is zero almost everywhere) and thus are useless.

We define the optimum probability threshold $p_{opt}$, as the probability threshold with minimum DST. The interested reader can find further details of the classifier quality in Appendix \ref{appendix:sensitivity}.

\subsubsection{Final classification} 
\label{subsection:final_classification}
The cluster and field models are then used to compute the cluster and field likelihoods for each source in our data sets. Missing values are marginalized over all their possible ranges with a uniform prior (see Assumption \ref{ass:MCAR}). In practice, the marginalization consists simply in restricting the representation space to the features actually present in the source. With the likelihoods and the fraction of members and non-members set as prior probabilities we compute the cluster posterior probabilities \cite[see Eq. 7 of][]{2014A&A...563A..45S}. Finally, we classify the sources as cluster members if their posterior probability is greater than the optimum probability threshold $p_{opt}$ derive above. 

\subsubsection{Posterior analysis} 
\label{subsection:posterior_analysis}

After our membership selection converged to a list of candidate members, we use the latter as starting point, and repeat the entire process. We find that our membership selection process is virtually insensitive to the initial list of candidates or representation spaces as long as the photometric features of the latter are chosen among those with higher importance. However and as expected, we also find that our methodology is sensitive to the $p_{in}$ value.  Appendix \ref{appendix:sensitivity} gives the results of the various sensitivity tests we performed, and a detailed justification for our choice of $p_{in}$=0.7 for both GDR2 and DANCe data sets and $p_{in}$=0.6 for the DANCe Central one.

Tables \ref{table:gaia_data_and_probabilities}, \ref{table:dance_data_and_probabilities}, and \ref{table:dance-central_data_and_probabilities}, available at the CDS, contain the cluster membership probabilities of the sources in the GDR2, DANCe and DANCe Central data sets, respectively. The tables provide the following information: identifiers (i.e. \textit{Gaia} source id, and DANCe id), equatorial coordinates (at epoch J2000.0), and cluster membership probabilities (obtained from the different $p_{in}$ values). In particular, Table \ref{table:dance_data_and_probabilities} also contains the DANCe astrometry and photometry described in Section \ref{subsection:dance_dataset}.

\section{Results}
\label{section:results}

The improved membership selection methodology introduced in the previous section provided us with a list of candidate members for each of our data sets. In this section we analyze these candidate members, compare the empirical isochrones they produce with theoretical ones, and derive their luminosity, mass and spatial distributions.

\subsection{Analysis of members}
\label{subsection:analysis_members}

In this section we analyze the lists of candidate members from the GDR2 (hereafter those of $p_{in}=0.7$), DANCe (hereafter those of $p_{in}=0.7$), and DANCe Central (hereafter those of $p_{in}=0.6$) data sets, give their general properties, compare them between themselves and with those from the literature. 

We notice that the candidate members resulting from the analysis of each data set are tightly linked to the learned classifier in that data set. Thus, in the following we use the same data set names when referring to the classifiers and its resulting candidate members.

\subsubsection{Comparison of candidate members}
\label{subsection:members_comparison}

Our three data sets have different origins, are made of different observables and have different properties in general, but a comparison is possible nevertheless. A good overlap between the three results would confirm the robustness of our analysis. We first cross-match the catalogues using a 1\arcsec\, search radius. The GDR2 data set has 4\,862\,245 sources in common with the DANCe data set, and 873\,500 with the DANCe Central one. The DANCe Central is by definition, fully contained within the DANCe data set.

\begin{figure*}
    \centering
    \includegraphics[width=\textwidth]{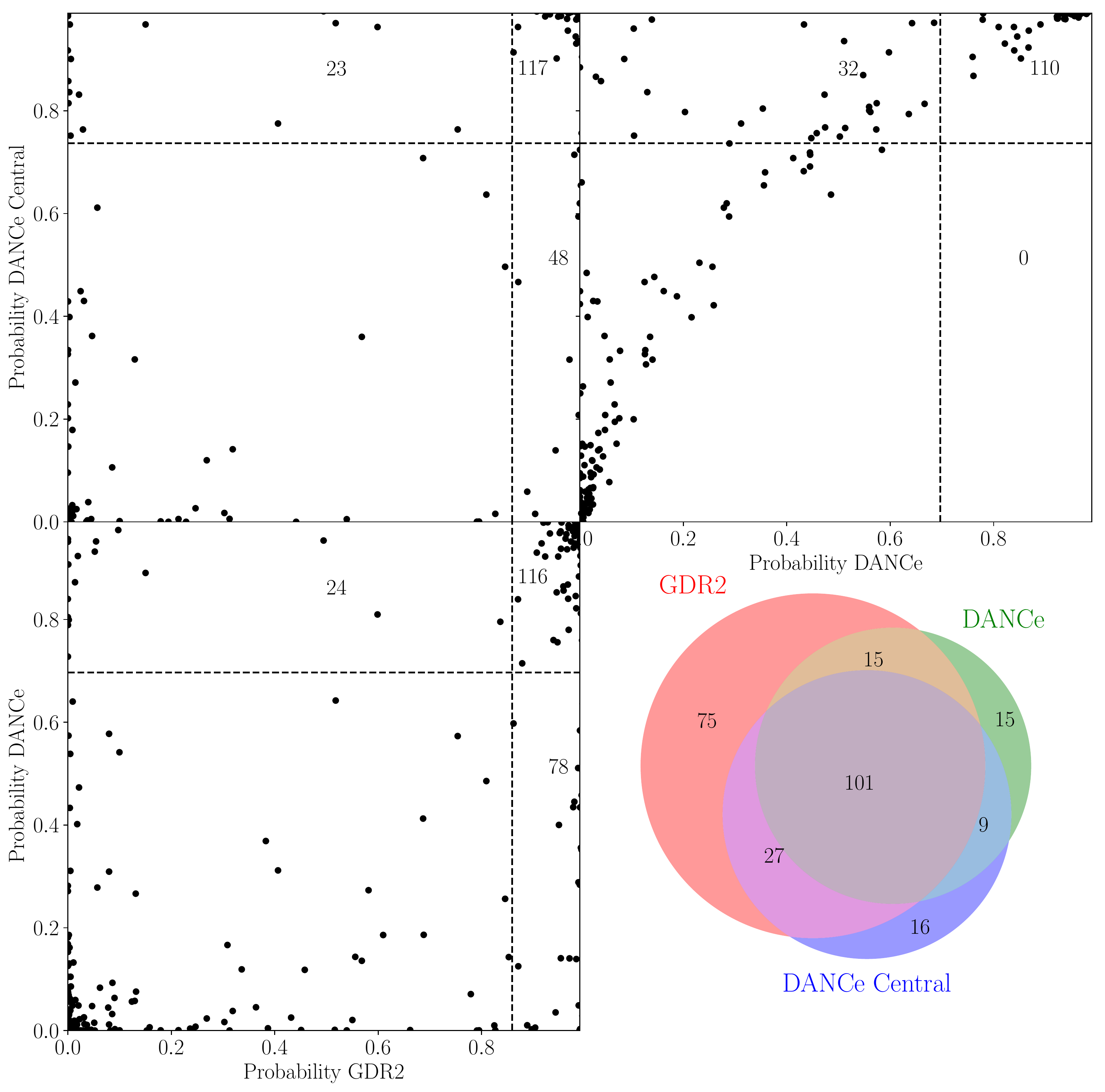}
    \caption{Comparison of candidate members from the three data sets. The upper panels and the bottom left one compare the membership probabilities of objects shared by the data sets. In all these panels, the horizontal and vertical dashed lines show the optimum probability thresholds. The bottom right panel shows a Venn diagram with all candidate members from the three data sets. We notice that the later also contains the GDR2 candidates outside the DANCe region and the WDs recovered from the $RS_2$, which are not included in the other three panels.  }
    \label{fig:prob_comparison}
\end{figure*}

In Fig. \ref{fig:prob_comparison} we compare the recovered membership probabilities of sources from the GDR2 ($RS_0$) data set versus those in the DANCe ($RS_1$, bottom left panel) and DANCe Central ($RS_1$, upper left panel) ones, and those from the DANCe data set versus those in the DANCe Central one (both in $RS_1$, upper right panel). In all these panels, the vertical and horizontal lines show the optimum probability thresholds ($p_{opt}$) of each data set. The number of accepted, rejected and common candidate members are shown within the corresponding boxes. In addition, the bottom right panel of Fig. \ref{fig:prob_comparison} also shows a Venn diagram with all candidate members from the three data sets (including the GDR2 candidates outside the DANCe region, and the WDs from $RS_2$).

As can be seen from Fig. \ref{fig:prob_comparison}, the majority of the candidate members are common to all data sets with minor differences in the classification. In the following paragraphs we discuss the similarity and differences of the three samples.

In the GDR2 vs. DANCe comparison, we observe 116 objects in common, with 24 DANCe candidates rejected by GDR2, and 78 GDR2 candidates rejected by DANCe. The proper motions of all rejected candidates (from both GDR2 and DANCe) are nonetheless compatible with those of the candidates in common. However, the majority (16 of 24) of the DANCe candidates rejected by GDR2 have parallaxes beyond the limits of the objects in common, which explains most of the discrepancies. On the other hand, the majority of the GDR2 candidates rejected by DANCe have photometry with either missing values (60 objects) in the representation space or/and values below 14~mag (28 objects), where our images start to saturate. {We notice that objects with missing values in the photometry have less evidence to overcome the larger field prior than the completely-observed sources, which results in lower membership probabilities. Although these objects are good candidate members, the reduction in the membership probability is enough to place them below the optimum probability threshold.}

In the GDR2 vs. DANCe Central comparison we observe a similar effect as in the DANCe case. The parallaxes of the majority (13 out of 23) of DANCe Central candidates rejected by the GDR2 classifier are beyond those of the candidates in common. However, the number of GDR2 objects rejected by DANCe Central is lower than in the DANCe classifier, only 48. From these latter, 34 have missing values in at least one band (of the representation space) and 19 report values lower than 14~mag (where detectors depart from linearity). This better performance of the DANCe Central classifier compared to that of the DANCe classifier, results from the better photometric completeness at the DANCe Central data set.

In the DANCe vs. DANCE Central comparison, we observe that the DANCe Central classifier gives higher membership probabilities with respect to the DANCe one. This effect results from the different size and extensions of the data sets: the DANCe one contains 3 times more sources, and in a region 8 times larger (see Section \ref{section:datasets}) than the DANCe Central one. Since we are adding the outer rings of the cluster, it is to be expected that the fraction of cluster members in the total data set decreases, or equivalently, the contamination by field sources in the regions of overlap in the representation space is higher. Hence, more restrictive probability thresholds are required to maintain similar contamination rates. The comparison shows that are 110 objects in common between these to classifiers. Also, there are 32 candidates from DANCe Central classifier rejected by the DANCe one. Nevertheless, these could be true cluster members rejected by the more restrictive DANCe classifier.

In the following we use as candidate members those resulting from the union of the three independent lists due to the lack of consistent evidence against the membership of sources rejected by one analysis but included by another. In Table \ref{table:members}, available entirely at CDS, we provide an extract of our 260 candidate members. 

\subsubsection{General properties}
Table \ref{table:statistics} shows a summary of the astrometric properties, radial velocities, and cluster membership probability ($\rm{p_{member}}$) of the candidate members recovered by each classifier. The values in this table are in agreement with those reported by \citet{Cantat-Gaudin2018}, \citet{Babusiaux2018} and \citet{2013AJ....145..134C}. We notice that the small shifts observed among lists are negligible when compared to the dispersion. As can be observed, there is an excellent agreement amongst the properties recovered from the three data sets.

\begin{table}[h]
    \centering
    \caption{Statistics of the candidates recovered in our three data sets.}
    \label{table:statistics}
    \resizebox{\columnwidth}{!}{
    \begin{tabular}{l|rr|rr|rr|}
\hline
\hline
{} &  \multicolumn{2}{c}{DANCe Central}  & \multicolumn{2}{|c|}{DANCe} &  \multicolumn{2}{c|}{GDR2} \\
{} & Mean & SD & Mean & SD & Mean & SD \\
\hline
\texttt{ra} (\degr)                    &               289.08 &              0.46 &       288.95 &      1.03 &      289.04 &     1.18 \\
\texttt{dec} (\degr)                   &               -16.33 &              0.57 &       -16.30 &      0.98 &      -16.45 &     1.57 \\
\texttt{pmra} (mas yr$^{-1}$)          &                -1.03 &              0.56 &        -0.95 &      0.47 &       -0.99 &     0.65 \\
\texttt{pmdec} (mas yr$^{-1}$)         &               -26.75 &              0.67 &       -26.63 &      0.66 &      -26.69 &     0.73 \\
\texttt{parallax} (mas)                &                 3.23 &              0.30 &         3.17 &      0.35 &        3.26 &     0.14 \\
\texttt{radial\_velocity} (km s$^{-1}$) &                43.02 &              6.60 &        43.56 &      7.32 &       40.02 &    16.65 \\
$\rm{p_{member}}$                      &                 0.96 &              0.07 &         0.93 &      0.07 &        0.98 &     0.03 \\
\hline
\end{tabular}

    }
\end{table}

\begin{figure}[ht!]
\begin{center}
\includegraphics[width =0.5\textwidth]{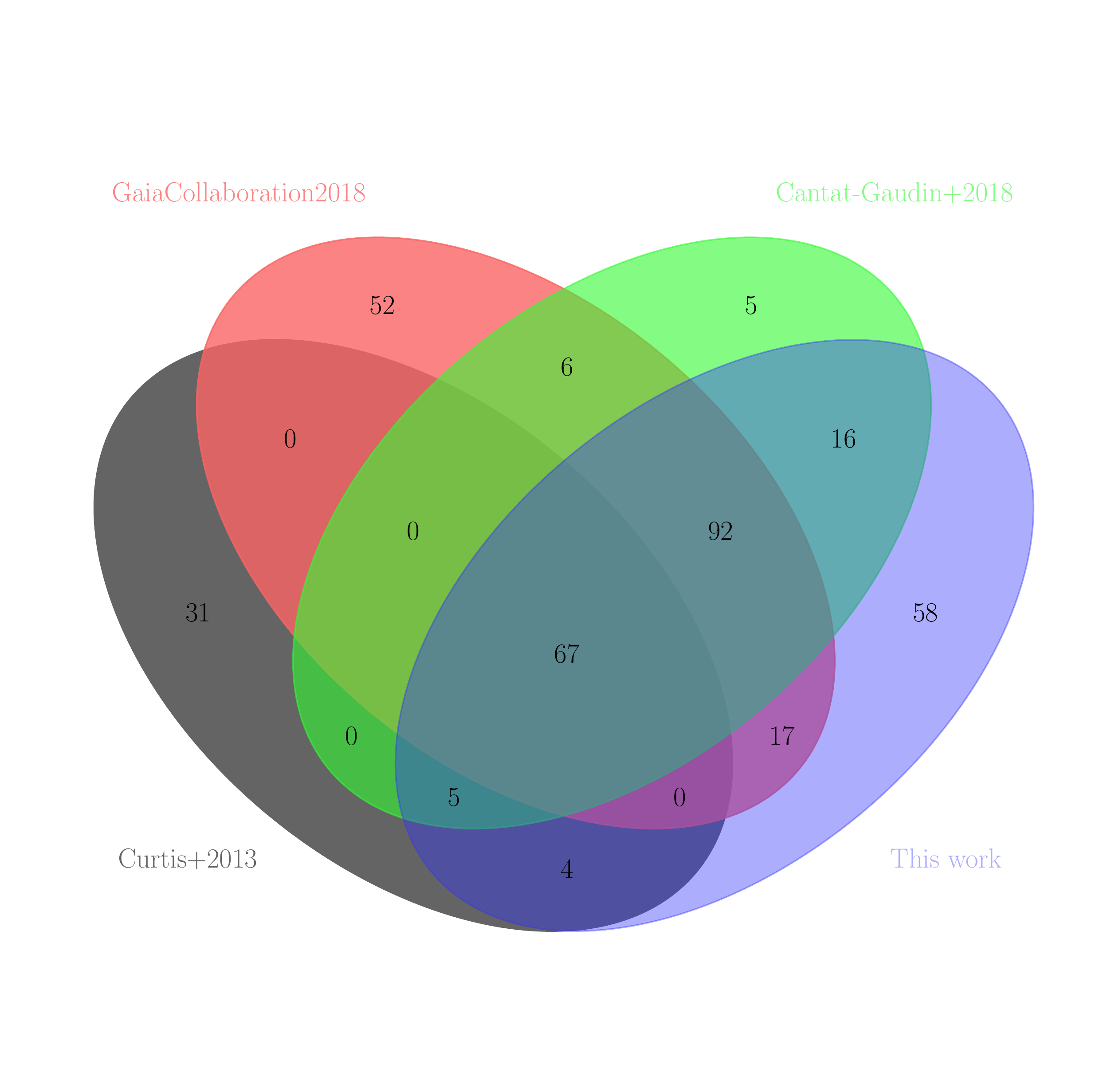}
\vspace{-1cm}
\caption{Venn diagram of candidate members of this work and from the literature.}
\label{fig:venn}
\end{center}
\end{figure}

\begin{figure*}
\begin{center}
\includegraphics[width = \textwidth]{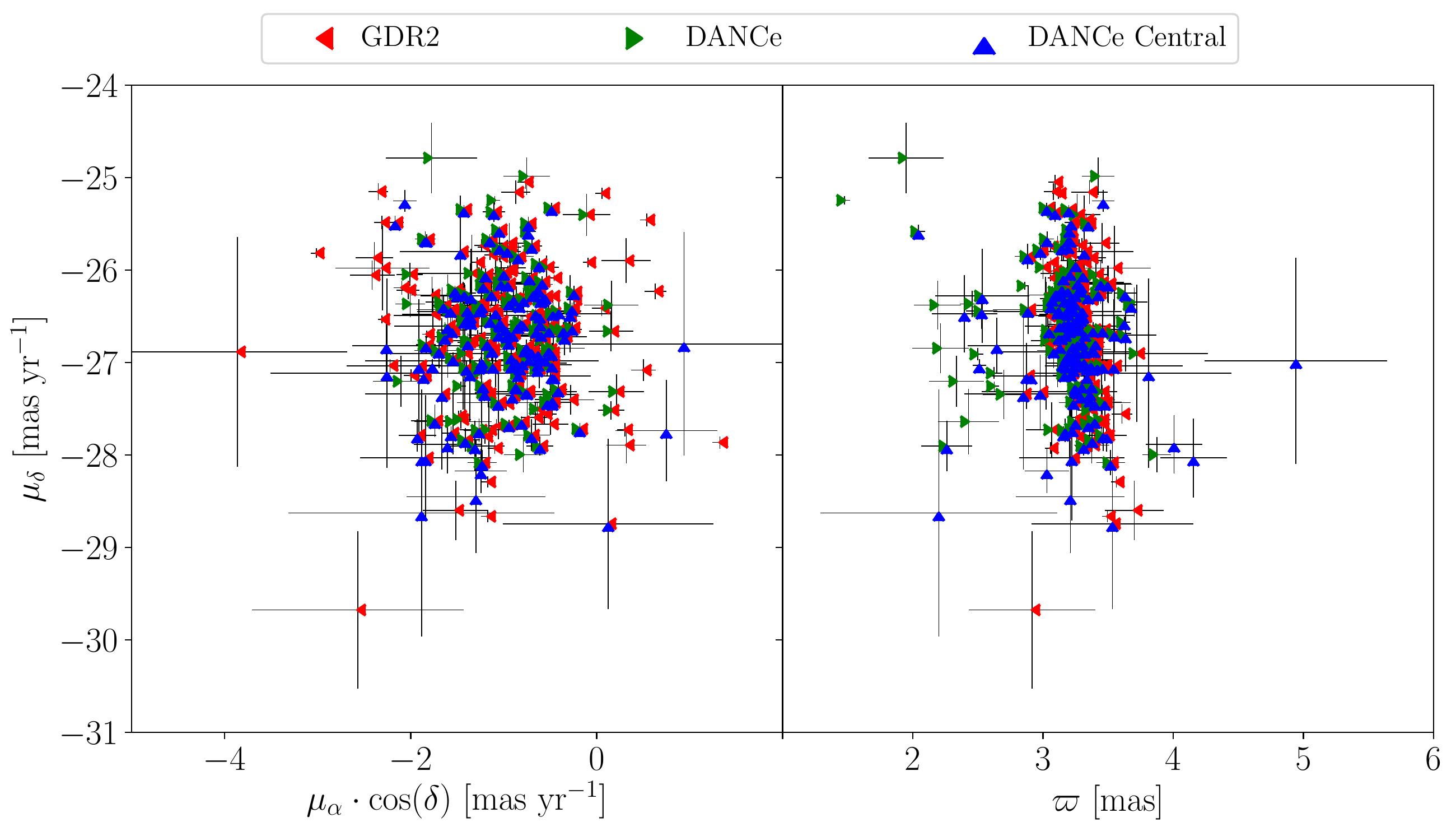}
\caption{Proper motions and parallaxes of our candidate members. The shape and colour indicate the classifier of origin.}
\label{fig:PPD}
\end{center}
\end{figure*}

Figure \ref{fig:PPD} shows the proper motions and parallaxes of all our candidate members. As can be seen, the DANCe and DANCe Central candidates show larger dispersion in the parallaxes, which is expected since this highly discriminant observable was not used to select these candidates. Nevertheless, all our candidate members are contained within $\pm2$ standard deviations of the DANCe mean parallax (see Table \ref{table:statistics}). However, using the mean and standard deviation of the GDR2 candidate members, there are 16 and 13 DANCe and DANCe Central candidate members, respectively, that lay beyond $\pm3$ standard deviations. In spite of this dispersion, the previous objects have proper motions similar to those of the GDR2 candidate members.

\begin{figure*}
\begin{center}
\includegraphics[width =\textwidth]{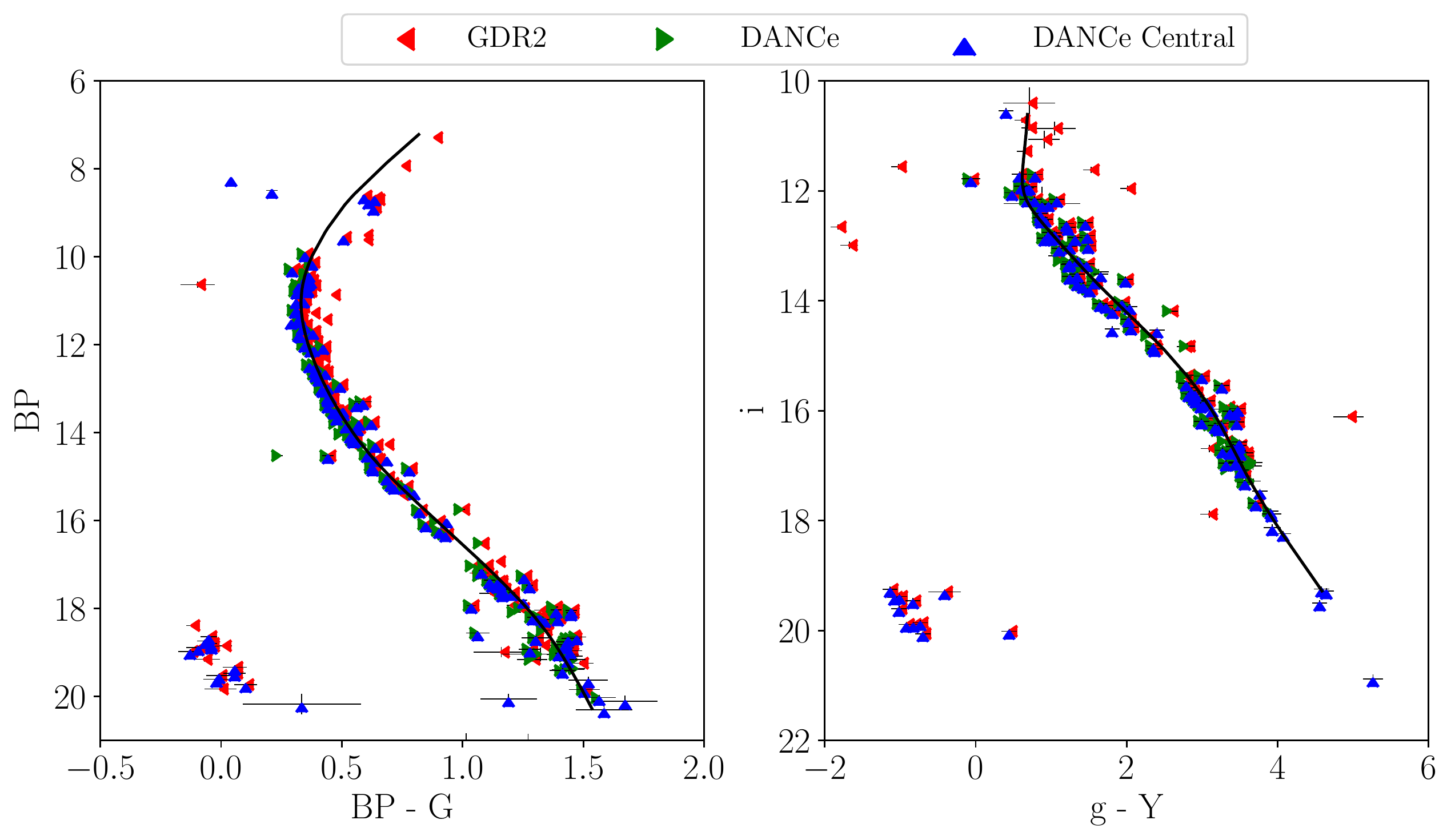}
\caption{Colour-magnitude diagrams of our candidate members in the \textit{Gaia} (left panel) and DANCe (right panel) photometry. The black lines show the empirical isochrones in the \textit{Gaia} and DANCe photometry.}
\label{fig:CMD}
\end{center}
\end{figure*}

Figure \ref{fig:CMD} shows the CMD of all our candidate members in the \textit{Gaia} and DANCe photometry. In addition, it also shows the projections of the multidimensional principal curve representing the cluster empirical isochrone. As can be seen in this figure, all our non-WDs candidate members are located in a clear sequence. Except those at the bright end, where the photometric detectors depart from linearity and report fainter magnitudes (objects with \texttt{i} < 14~mag and \texttt{g-z} $\sim2$). On the other hand, our WD candidates are grouped in the cooling sequence at the bottom left of the CMDs. 

Tables \ref{table:isochrone_gaia} and \ref{table:isochrone_dance}, available at the CDS, provide apparent \textit{Gaia} and DANCe magnitudes, respectively, of the cluster empirical isochrones. The latter were computed excluding WDs and objects with missing values.

According to the radial velocity statistics reported in Table \ref{table:statistics}, three candidate members have radial velocities beyond 3 standard deviations of the mean. Two of them are located above the cluster sequence in the CMD, and near the equal-mass binary sequence, which make them good binary candidates. Nonetheless, the three have parallaxes compatible with the clusters (less than 2 standard deviations from the mean). Thus, we decide to keep these three objects in our list of candidate members given their low number, high cluster membership probability ($>0.9$), and lack of evidence to reject them as cluster members.

\subsubsection{Comparison with the literature}

We now compare our list of candidate members with those reported by \citet{Cantat-Gaudin2018}, \citet{Babusiaux2018}, and \citet{2013AJ....145..134C}. We cross-matched our candidates with those of the previous authors using \textit{TOPCAT} \citep{TOPCAT} and a maximum allowed separation of 1\arcsec. We find the common and rejected objects shown in Fig. \ref{fig:venn}.

From the list of astrometric candidate members of \citet{Babusiaux2018} we recover 176 and reject 58. However, only 4 are truly rejected since the remaining ones do not pass the astrometric (parallax over error >10) and photometric (magnitude over error > 20) filters applied by those authors.

From the list of candidate members of \citet{Cantat-Gaudin2018} we recover 180 and reject 11. The rejected sources have cluster membership probabilities in the range 0.2 to 0.8, which still make them probable candidate members. Nonetheless, they are below our optimum probability threshold.

From the list of candidate members of  \citet{2013AJ....145..134C} we recover 77 and reject 31, although one of them is not present in \textit{Gaia}, see Section \ref{subsection:previous_members}. From the latter, 5 have discrepant parallaxes (beyond 20 standard deviations of the cluster mean GDR2 parallax), 5 have discrepant proper motions (beyond 5 standard deviations of the mean GDR2 proper motions), and 5 have discrepant photometry (away from the cluster sequence). From the remaining, 10 have large uncertainties, which give them low membership probabilities (<0.1), and only 5 are probable candidate members that lay below our optimum probability threshold. 

From the previous comparisons we notice that:
\begin{itemize}
\item the list of candidate members by \citet{Cantat-Gaudin2018} is the one with the largest number of common members, which is expected given the the similarities of both methodologies.
\item the number of probable members rejected by our methodology (4, 11 and 5 from the list of \citealt{Babusiaux2018}, \citealt{Cantat-Gaudin2018}, and  \citealt{2013AJ....145..134C}, respectively) is well contained within the poisson statistics, and thus for our purposes, there is no need to include them in our list of candidates.
\end{itemize}

\subsubsection*{Radial velocities of rejected candidate members}
Radial velocity is commonly used as an indicator for cluster membership. However, it is not used in our membership selection due to the large number of objects in which  it is missing. Furthermore, it is also independent of the \textit{Gaia} astrometric solution. Therefore, the radial velocities of the candidate members from the literature that were rejected as such by our analysis offer an excellent opportunity to study the quality of our data set and membership selection method.

None of the candidate members of \citet{Babusiaux2018} rejected in our analysis have \textit{Gaia} radial velocity measurements.

Only one of the candidate members of \citet{Cantat-Gaudin2018} rejected by our analysis (\textit{Gaia} ID 4084598290324148480) has a \textit{Gaia} radial velocity measurement. The latter is further away than 4 standard deviations of the cluster mean (see Table \ref{table:statistics}), which together with its moderate cluster membership probability ($p=0.66$) make it a probable binary member.

Amongst the candidate members of \citet{2013AJ....145..134C} that were rejected by our analysis, 28 have radial velocity measurements (19 from \textit{Gaia} and the 28 from the mentioned authors), and all these are compatible with the cluster (within 2 standard deviations of the mean, see Table \ref{table:statistics}). Four of these objects (CWW 5, 12, 16, 24), were classified by \citet{2013AJ....145..134C} as Blue Stragglers. We notice that the latter is a population not fully modelled by our methodology (see Assumption \ref{ass:populations}). In addition, seven objects (CWW 18, 24, 35, 62, 72, 80 and 84) have non-negligible membership probabilities ($0.1 < p < 0.8$), which make them probable members. Four of these (CWW 35, 62, 72 and 80) have large \texttt{astrometric\_excess\_noise} (> 1.0), pointing to possible binarity, which is the case of the double-line spectroscopic binary CW 72 \citep{2013AJ....145..134C}. Finally, \citet{2013AJ....145..134C} mentioned that CWW 51 is a double star separated by 1.65\arcsec, with both components having radial velocities compatible with the cluster \citep{2016PhDT.......246C}. Nonetheless, our methodology reports different membership probabilities for these objects, while  $ID_{\rm{member}}$=221 has a $p=0.999$, the companion (Gaia ID 4184134810243354368) has a low membership probability, $p=0.06$.

Most likely the binarity of these objects is impacting its astrometric and photometric measurements in ways that our membership selection is currently unable to deal with. Thus, from the previous analysis we notice the that:
\begin{itemize}
    \item The population of binaries is probably underestimated. As pointed out by \cite{2018A&A...616A...2L}, resolved or unresolved binaries may produce spurious astrometric results in the \textit{Gaia} data.
    \item Our methodology misses cluster members in late stellar evolution stages, e.g. Blue Stragglers.
\end{itemize}

\subsubsection*{New members}
Our methodology finds 58 new candidate members, which represents an increase of 30\% in the number members reported in the literature. From these, 28, 14 and 8 originate only from GDR2, DANCe and DANCe Central classifiers, respectively, the remaining 8 are shared by the three classifiers. From the 28 members only present in the GDR2 list, 25 are located in the cluster out-skirts beyond the region analyzed by \citet{Cantat-Gaudin2018} and \citet{Babusiaux2018}. The parallaxes of  all the new GDR2 candidate members are located within 5 standard deviations of the previously known candidate members from the literature. However, the parallaxes of 10 of the 23 new candidate members from the DANCe and DANCe Central classifiers lay beyond the previous limit. Although these objects are spread in parallax (up to 15 standard deviations), their proper motions and photometry are still compatible with the cluster sequence, which give them enough membership probability to classify them as cluster members. 

In addition, the magnitude distribution of our GDR2 candidate members agrees with those obtained from the candidate members of  \citet{Cantat-Gaudin2018} and \citet{Babusiaux2018}, see Fig. \ref{fig:magnitude_distributions}. Nevertheless, we notice the large contamination rate, $\sim$72\%, present in the astrometric candidate members of \citet{Babusiaux2018} at the faint magnitudes (>18 mag).

\subsection{Empirical isochrone}
\label{subsection:isochrone}

In this section we compare the empirical isochrone given by our candidate members with the theoretical ones of  COLIBRI \cite[][]{Marigo2017}\footnote{\label{footnote:cmd}PARSEC v1.2S + COLIBRI PR16 model downloaded from \url{http://stev.oapd.inaf.it/cgi-bin/cmd}.}, MIST \citep{2016ApJS..222....8D,2016ApJ...823..102C}\footnote{\url{http://waps.cfa.harvard.edu/MIST/index.html}}, and BT-Settl \citep{2014IAUS..299..271A}. To perform this comparison, the distance to the stars and its extinction are needed. While the \textit{Gaia} DR2 data provides individual parallaxes and $A_G$ extinctions, the DANCe data does not. Thus, we perform two comparisons. In the first one, we use the \textit{Gaia} photometry, parallax, and extinction to obtain absolute magnitudes. In the second one, we use the typical cluster distance and extinction, obtained in the previous comparison, and derive absolute magnitudes for the DANCe photometry. 

\subsubsection{Gaia photometry}

To derive the absolute \textit{Gaia} magnitudes of our candidate members we use their individual parallax and its uncertainty. We compute distances by means of the \emph{Kalkayotl}\footnote{\url{https://github.com/olivares-j/kalkayotl}} code (Olivares et al. in prep.). In a nutshell, it infers the posterior distribution of the distance to the star given its parallax distribution (mean and uncertainty) and a distance prior specifically tuned for stars in open clusters. We tested different distance priors (uniform, Gaussian, Cauchy and Exponentially Decreasing Space Density, \citealt{2015PASP..127..994B}) in synthetic data sets with properties mimicking the real one. We find that the Cauchy one reproduces better the synthetic true distances. 

\begin{figure}[htp]
\begin{center}
\includegraphics[width = \columnwidth]{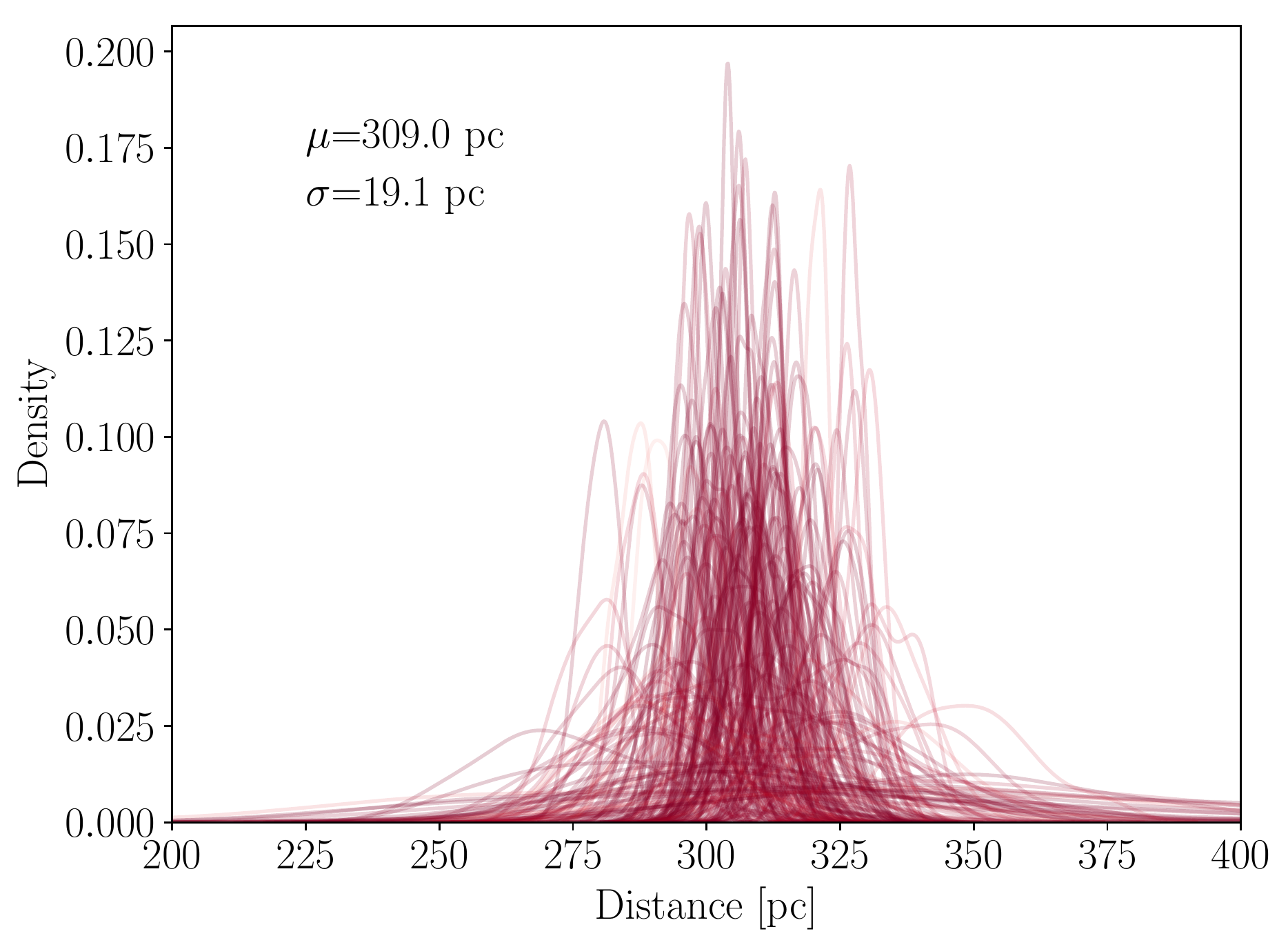}
\caption{Distance distribution of the GDR2 candidate members. Each line represents the KDE of samples from the posterior distribution of individual distances.}
\label{fig:distances}
\end{center}
\end{figure}

Figure \ref{fig:distances} shows the distribution of distances to our GDR2 candidate members, the mean and standard deviations are $\mu=309$ pc and $\sigma=19$ pc. This mean distance value is compatible within the uncertainties with that obtained by simple inversion of the mean parallax ($\mu_{\omega} = 3.2$ mas), which results in a typical distance of 312 pc.

Then, we compute absolute magnitudes as follows. First, we draw 100 samples from the uncertainty distribution(see Assumption \ref{ass:gaussian_uncertainties}). Then, each of the previous samples is transformed to absolute magnitude using the 100 samples of the distance distribution. Afterwards, we compute the 2.5, 50 and 97.5 percentiles of the resulting absolute magnitude distributions.

\begin{figure}[htp]
\begin{center}
\includegraphics[width = \columnwidth]{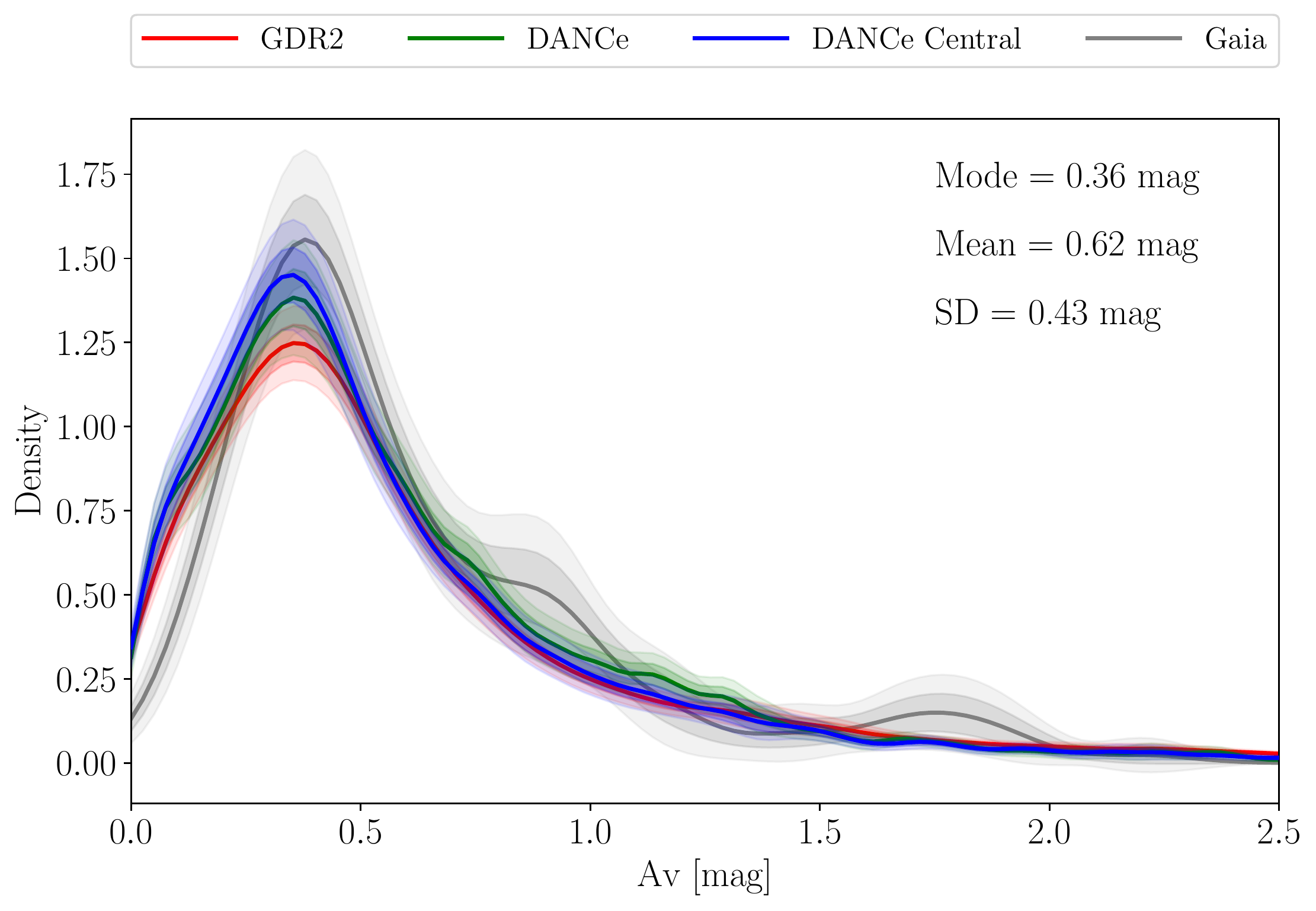}
\caption{Density of $A_V$ values for our candidate members (solid lines) and their 1$\sigma$ and $2\sigma$ uncertainties (shaded regions, computed from 100 bootstrap samples). Colour codded are the densities resulting from the \textit{Gaia} DR2 values (grey lines and reported statistics), and those inferred from our GDR2, DANCe and DANCe Central candidates (red, green and blue, respectively), see Section \ref{subsection:extinction}.}
\label{fig:extinction}
\end{center}
\end{figure}

Concerning the cluster extinction, \citet{2013AJ....145..134C} derived a value of $A_V = 0.25 \pm 0.05$ mag. However, these authors mention that due to cloud structure, $A_V$ varies between 0.3 and 0.5 mag. Since \textit{Gaia} DR2 also provides extinction in the \texttt{G} band for some stars, we analyze the transformed extinction distribution in the $V$ band. We transformed the $A_G$ values to $A_V$, using the extinction coefficient  $A_G/A_V = 0.85926$ (as provided by the CMD simulator, see note \ref{footnote:cmd}). Figure \ref{fig:extinction} shows the KDE of the $A_V$ of our candidate members (grey line). As can be seen, the distribution is highly asymmetric with a clear peak at $A_V=0.36$ mag, and some objects reaching 2~mag of extinction. Although we have individual $A_V$ values, we decided to use the ensemble typical value, as recommended by \citet{2018A&A...616A...8A}. The previous Fig. also shows the KDE of the inferred extinction values (see Section \ref{subsection:extinction}), which are all compatible with those reported by \textit{Gaia} DR2.

The left panel of Figure \ref{fig:absolute_CMD} compares the MIST and COLIBRI theoretical isochrones to all our candidate members with \textit{Gaia} photometry. Unfortunately, at present, the BT-Settl model includes only the pre-launch \textit{Gaia} photometry, and thus we cannot use it. 

It is clear from Figure \ref{fig:absolute_CMD} that there is a good agreement between the empirical and theoretical isochrones of both models. However, there are some objects located far from the theoretical isochrones, the majority of them coming from the DANCe Central classifier. We remind the reader that the highly discriminant parallax feature was not used in the classification of the DANCe and DANCe Central objects. Although these might seem contaminants, their discrepant photometry can also be caused by problems in the \texttt{BP} photometry (see below).

\subsubsection{DANCe photometry}
\begin{figure*}[htp]
\begin{center}
\includegraphics[width = \textwidth]{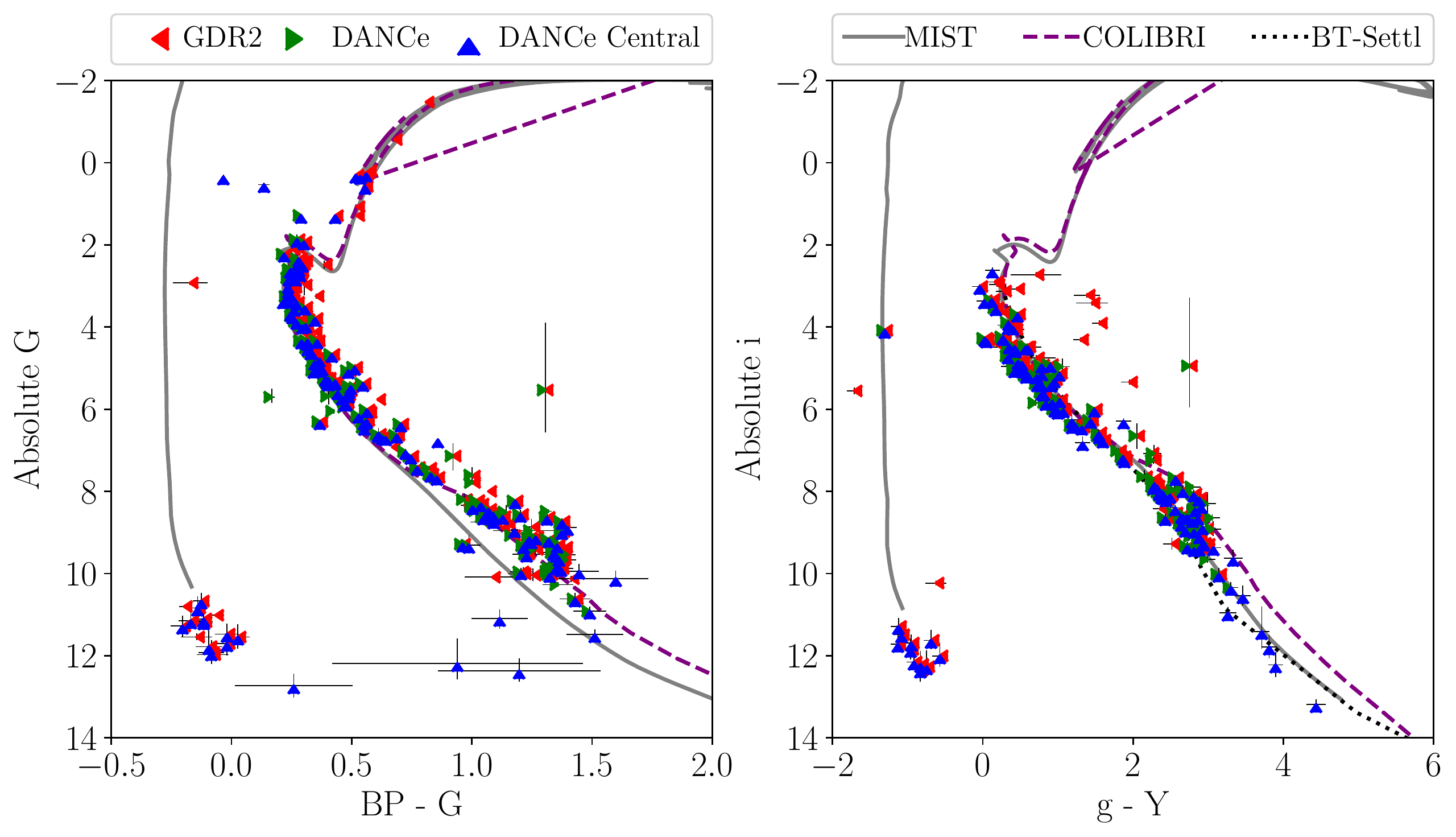}
\caption{Colour-magnitude diagrams showing the absolute magnitude of our candidate members in the \textit{Gaia} photometry (left panel) and DANCe photometry (right panel). Also shown are the COLIBRI, MIST and BT-Settl isochrones.}
\label{fig:absolute_CMD}
\end{center}
\end{figure*}

The DANCe data set does not provide parallaxes nor extinction values for any object beyond the \textit{Gaia} limit. Thus, for those objects with no parallax we use the typical distance (309 pc and 19 pc of dispersion) found in the previous section. For all objects we compute absolute magnitudes as described in the previous section. The right panel of Figure \ref{fig:absolute_CMD} shows the absolute \texttt{i} vs \texttt{g-Y} colour-magnitude diagram of all our candidate members, together with the 2.5 Gyr COLIBRI, MIST and 2.0 Gyr BT-Settl isochrones. As can be seen, the agreement between empirical and theoretical isochrones is still good.

From the comparison of the right and left panels of Figure \ref{fig:absolute_CMD}, we observe the following aspects. First, those objects located far from the theoretical isochrones at faint \textit{Gaia} magnitudes are closer to the isochrones in the DANCe photometry, which may be indicative of problems in the \texttt{BP} photometry of these red and faint objects. Moreover, the \texttt{BP} uncertainties of these red objects might be underestimated. Second, the very bright objects in the \textit{Gaia} photometry, those approching the red clump, are displaced to fainter magnitudes in the DANCe photometry. This effect results from the non-linearity of the photometric detectors at this bright region.

As shown in this section, the empirical and theoretical isochrones are in good agreement, at the bright regions particularly. Nevertheless, the COLIBRI isochrones provide a better agreement at absolute magnitudes \texttt{G} and \texttt{i} > 8 mag. Since the BT-Settl isochrone includes only the pre-launch \textit{Gaia} photometry, and the MIST ones provide a poorer fit at faintest magnitudes, in the following we will work only with the COLIBRI ones.

\subsection{Luminosity distribution}
\label{subsection:luminosity}

In this Section our objective is to derive the luminosity distribution of the cluster. Since our list of candidate members comes from diverse origins and have different observables and degrees of completeness, we analyze the luminosity distribution in three regions of completeness. The Core region covers the inner $3^{\circ}\times 3^{\circ}$ area, the Middle region covers the inner $8^{\circ}\times 6^{\circ}$ area, and the Full region covers the entire $6^{\circ}$ radius area.

We notice that it is impossible to disentangle the results obtained at each data set (classifier) from those of the region since each data set is associated to a specific region. Nevertheless, to make a fare comparison, we restrict the candidate members from the DANCe and GDR2 classifiers to the Core region, and those from the GDR2 classifier to the Middle region.

We have obtained the luminosity of each candidate member from two methods. The first method transforms samples of the absolute magnitudes (obtained in the previous section) to luminosities using the magnitude-to-luminosity relation (one for each photometric band) provided by the theoretical isochrone. To properly transform absolute magnitudes into luminosities, the transformation must be injective (i.e. a one-to-one function). However, the theoretical isochrones at this old cluster age show a loop in luminosity at the turn-off region. This loop prevents the transformation to be injective. Thus, we filter out this loop and smooth the magnitude-to-luminosity relation. The second method uses the \textit{Sakam}\footnote{\url{https://github.com/olivares-j/Sakam}} code to infer the posterior distribution of the candidate luminosity given its available absolute photometry, and the theoretical isochrone. 

We observe no difference between the luminosity distribution obtained from both methods within the completeness interval. However, at the bright end and outside the completeness limits, we observe a spurious abundance of bright sources in the luminosity resulting from the first method as a result of the smoothing of the magnitude-to-luminosity relation. Therefore, in the following we report only the luminosity distribution obtained from the second method.

\begin{figure*}[ht!]
\includegraphics[width =\textwidth]{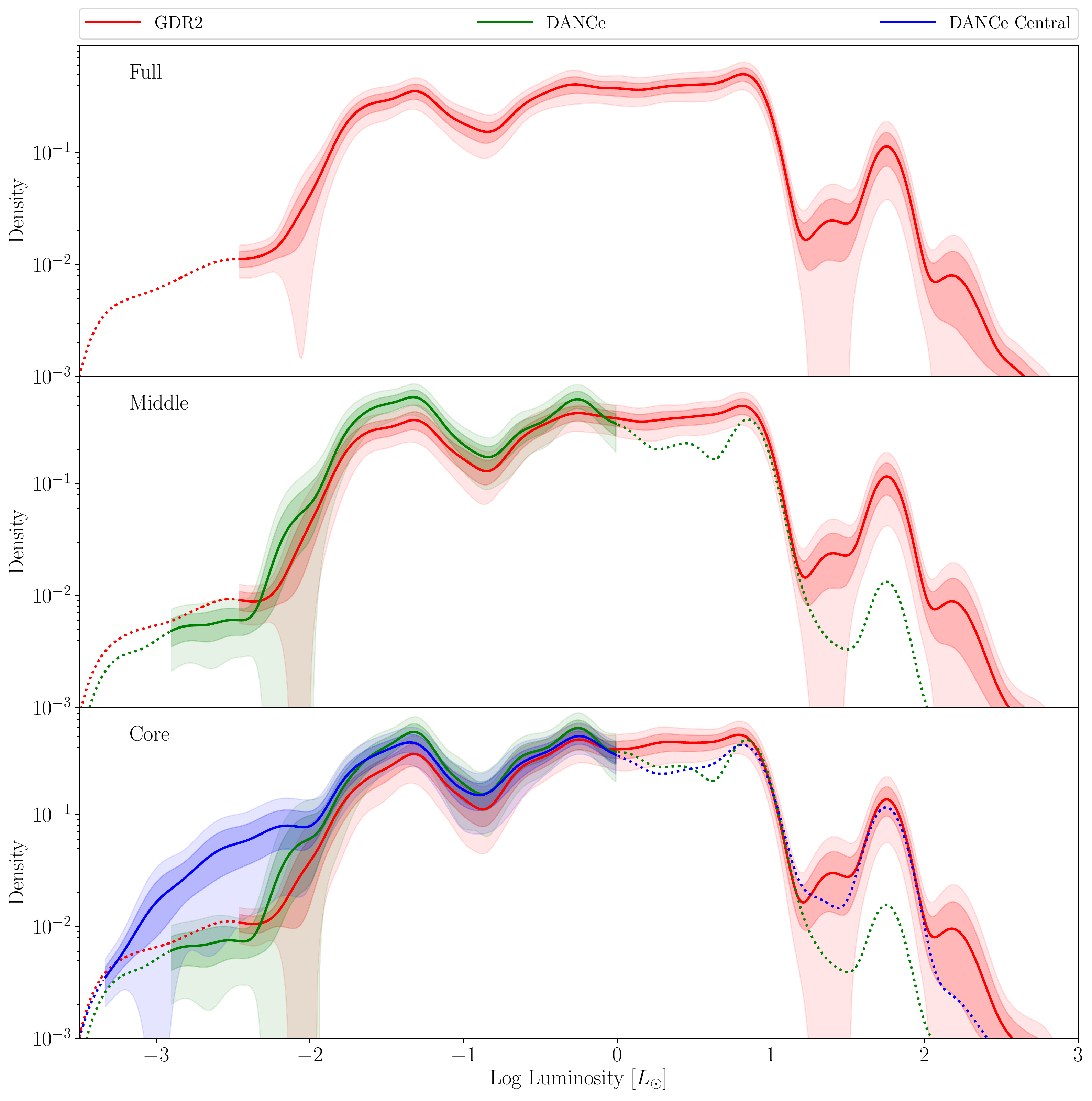}
\caption{Luminosity distributions resulting from the candidate members recovered by the GDR2, DANCe and DANCe Central classifiers in the three analyzed regions. The incomplete regions are represented with dotted lines. The 1$\sigma$ and 2$\sigma$ uncertainties (shaded regions) are computed from 100 bootstrap samples.}
\label{fig:luminosity}
\end{figure*}

Figure \ref{fig:luminosity} shows the luminosity distribution derived from the GDR2, DANCe, and DANCe Central candidate members in the Core, Middle and Full regions. To compute the luminosity of each source we use all its photometric bands, thus we use the minimum and maximum incompleteness limits of all bands to represent the the faint and bright incompleteness limits, respectively.

Comparing the panels of Fig. \ref{fig:luminosity}, we observe the following aspects. First, the differences in the luminosity distributions resulting from candidate members of the three classifiers are significative only at the bright and faint ends. At the bright end, there is a remarkable agreement between the results of the DANCe Central and GDR2 classifiers, even where the former is incomplete. We interpret this agreement as a product of the homogeneity of their spatial coverage, which is not the case in the DANCe classifier. On the other hand, at the low-luminosity end ($\log L/L_{\odot} < -2.0$) the DANCe  and DANCe Central classifiers report the lowest and highest density, respectively. The GDR2 classifiers reports similar densities as the DANCe one within its completeness limits. Although the differences in the reported densities at this region can be reconciled by the uncertainties (within the 2-$\sigma$ limits), it is clear that the DANCe Central classifier reports the highest density. The latter results from the four faintest candidate members that come exclusively from this classifier. 

Second, we observe signs of luminosity segregation. In the Core region, the three classifiers show a peak at $\log L/L_{\odot}\sim-1.3$. However, in the Middle and Full regions, a secondary peak is present at $\log L/L_{\odot}\sim-1.7$ giving rise to a plateau at $-1.7<\log L/L_{\odot}<-1.3$. Thus, the Core region contains less objects at these faint luminosities than the Middle and Full regions. In addition, the peak at $\log L/L_{\odot}\sim-0.2$ in the Core region is not present in the Full region. Thus indicating that there is an over-abundance of these intermediate-luminosity objects in the Core region.

In spite of the aforementioned differences, the following features in the luminosity distribution are common to all classifiers and analyzed regions. These features are, from bright to faint: i) a peak at $\log L/L_{\odot} \sim 1.8$, which corresponds to the red clump ( absolute \texttt{G} magnitude $\sim$ 0 in Fig. \ref{fig:absolute_CMD}), ii) a valley in the region: $1 < \log L/L_{\odot} < 1.5$, iii) a plateau from $-0.5 < \log L/L_{\odot} < 1$, iv) a dip at $\log L/L_{\odot} \sim -0.9$ covering the interval $-1.3 < \log L/L_{\odot} < -0.5$, and v), a pronounced drop at  $\log L/L_{\odot} \sim -2$. 

We notice that the dip at $\log L/L_{\odot} \sim -0.9$ has its counterpart in the magnitude distributions at \texttt{i}$\sim$ 14.5 mag, and \texttt{BP}$\sim$16~mag (see Fig. \ref{fig:magnitude_distributions}). It is consistently observed in spite of the representation space and $p_{in}$ value used, and is also present in the magnitude distribution of the candidate members of \citet{Cantat-Gaudin2018} and \citet{Babusiaux2018} (see Fig. \ref{fig:magnitude_distributions}). Furthermore, it is statisticaly significative beyond the 2-$\sigma$ level as can be observed in Fig. \ref{fig:luminosity}.

We hypothesize that this dip in the luminosity distribution has three possible explanations. It could be caused by a bias in the membership selection, an incompleteness in the data, or a true physical phenomenon. 

We can discard the hypothesis of a bias in the membership selection since the dip is also present in the magnitude distributions of the candidate members  of \citet{Cantat-Gaudin2018} and \citet{Babusiaux2018}, who use different methodologies. 

The dip appears well within the completeness limits of both the \textit{Gaia} and DANCe data, and as far as we know, there is no \textit{Gaia} incompleteness at this magnitude range (15~mag< \texttt{G} < 17 mag). Furthermore, it appears on both the GDR2 and DANCe sets, in spite of their different origins. Therefore, we also discard the hypothesis of incompleteness in the data.

Finally, our third possible explanation is a physical origin. Indeed, the Wielen dip \citep{1981AJ.....86.1898U}, located at $0.7M_{\odot}$ \citep{1990MNRAS.244...76K}, corresponds to a luminosity of $\log L/L_{\odot} \sim -0.9$. This dip has been observed in other young and intermediate age open clusters \cite[see for example][]{1997JKAS...30..181L,2001A&A...375..863J,2002MNRAS.335..291N}, and as proposed by several authors \cite[e.g.][]{1990MNRAS.244...76K,2002MNRAS.335..291N} it may result from a change in slope of the mass-luminosity relation. \citet{1990MNRAS.244...76K} argue that this change in slope is a consequence of an increasing importance of H$^-$ as a source of opacity. 

To the best of our knowledge, this is the first time that the Wielen dip has been confirmed in such an old open cluster.

\subsubsection{Inferred extinction}
\label{subsection:extinction}

In addition to the inference of each candidate member luminosity, the \textit{Sakam} code allows us to infer its extinction. Figure \ref{fig:extinction} shows the KDE of posterior samples of the inferred $A_V$ from our candidate members in the three data sets. As can be seen, our extinction values are compatible with those  reported by \textit{Gaia} DR2, and in excellent agreement with the value reported by \citet[][$0.347\pm0.09$ mag]{2018ApJ...866...67T} for the cluster member \object{EPIC 219394517}. The large dispersion of the $A_V$ values is a clear evidence of differential extinction, as originally reported by \citet{2013AJ....145..134C}.

\subsection{Mass distribution}
\label{subsection:mass}

\begin{figure}[htp]
\begin{center}
\includegraphics[width =\columnwidth]{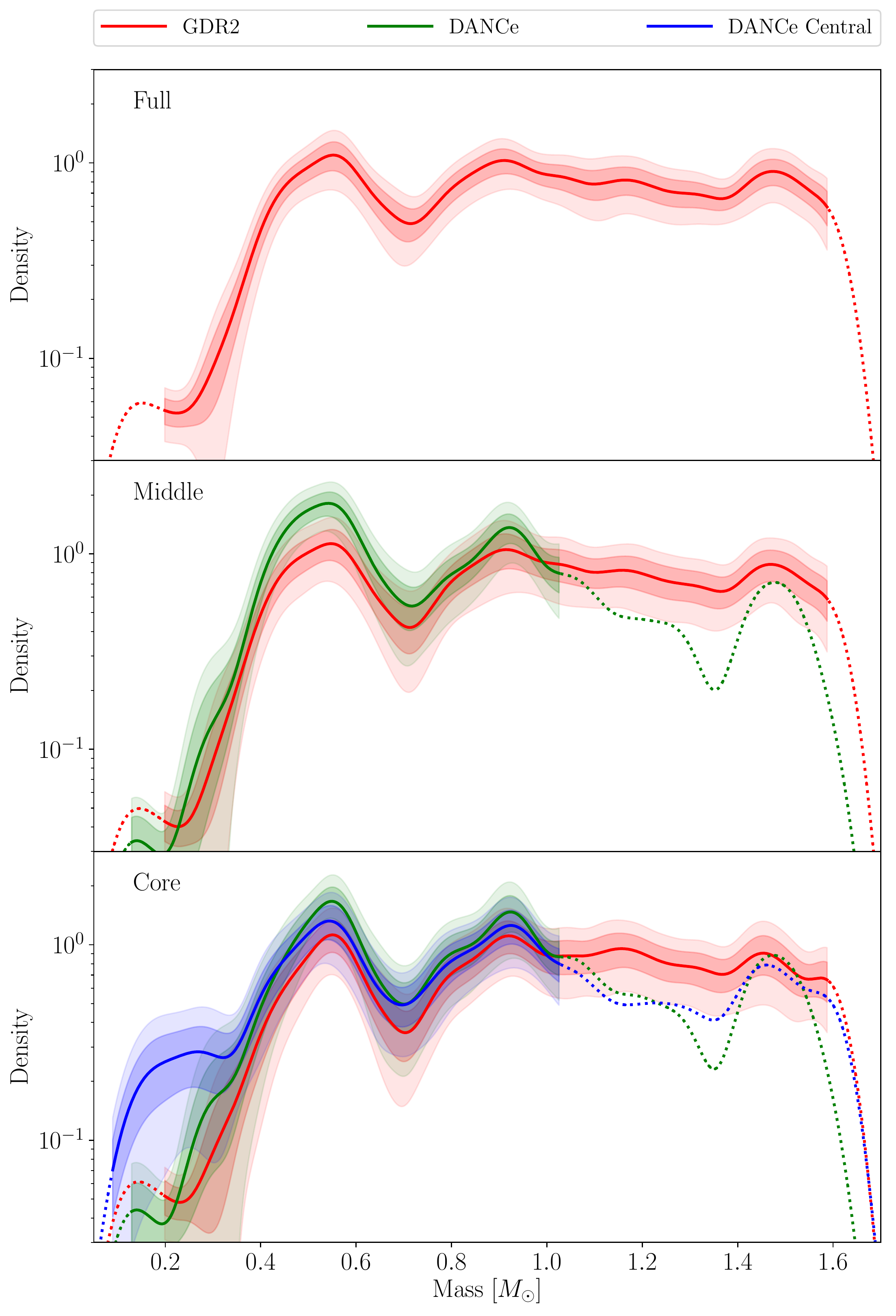}
\caption{Mass distribution of our candidate members inferred from the three data set of origin. The uncertainty  is represented with shaded regions showing 1$\sigma$ and 2$\sigma$ values obtained from one hundred bootstrap samples. Dotted lines indicate regions of incompleteness.}
\label{fig:mass}
\end{center}
\end{figure}

We derive the mass distribution of all our candidate members using the the \textit{Sakam} code. As in the previous section, it uses the theoretical COLIBRI isochrone and the available absolute photometry of each candidate member. However, we now infer the posterior distribution of its initial mass\footnote{The theoretical isochrones provide the initial (i.e. zero-age main sequence), and the present-day masses.}. We notice that this mass corresponds to the system mass since we are not able to disentangle unresolved binaries or multiple systems. Figure \ref{fig:mass} shows the KDE of all samples from the posterior mass distribution of our DANCe Central, DANCe and GDR2 candidates member (WDs excluded). The completeness limits correspond to those of the luminosity transformed into masses with the mass-luminosity relation.

In the previous figure, we observe a plateau from 0.4-1.6$M_{\odot}$, with a shortage of stars in the interval 0.6-0.8$M_{\odot}$ that corresponds to the Wielen dip, as previously discussed. We also observe that the mass distribution below 0.4$M_{\odot}$ has two different behaviours. On one hand, the mass distribution resulting from the DANCe Central classifier has a depression at $\sim$0.35$M_{\odot}$ followed by an almost flat plateau until 0.15$M_{\odot}$, and a fall at 0.1$M_{\odot}$, which marks the end of the isochrone. We notice that the depression at $\sim$0.35$M_{\odot}$ may corresponds to the Kroupa dip \citep{1990MNRAS.244...76K}. However, given the large uncertainties, we cannot statistically confirm it. On the other hand, the mass distributions resulting from the GDR2 and DANCe classifiers show an steady drop from 0.5-0.2$M_{\odot}$ and then a plateau from 0.2-0.1$M_{\odot}$. 

The over-density shown by the DANCe Central mass distribution with respect to to those of the DANCe and GDR2 results from the four faintest candidate members (see Fig. \ref{fig:CMD}), which were recovered only in this data set. Given the faint magnitudes (\texttt{i} > 19 mag) of these four objects, their membership should be revisited in the light of future observations.

In the high-mass range, the mass distributuion is limited by the cluster age. Indeed, stars with masses heavier than $\sim 1.6 M_{\odot}$ might have already evolved into WDs or supernovae.

In order to extend the mass distribution beyond the limit imposed by the cluster age, we complement the inferred mass distribution with the masses of our WDs. We cross-match our 15 WDs candidates with the catalogue of \citet{2018arXiv180703315G} and found 14 matches.  We transform the final hydrogen masses \citep[provided by the catalogue of][]{2018arXiv180703315G} to initial masses using the empirical WD initial-to-final mass relation (IFMR) of \citet{2018ApJ...866...21C}. For the object with no mass estimate we use a uniform distribution in the range of the IFMR.

\begin{figure*}
\begin{center}
\includegraphics[width = \textwidth]{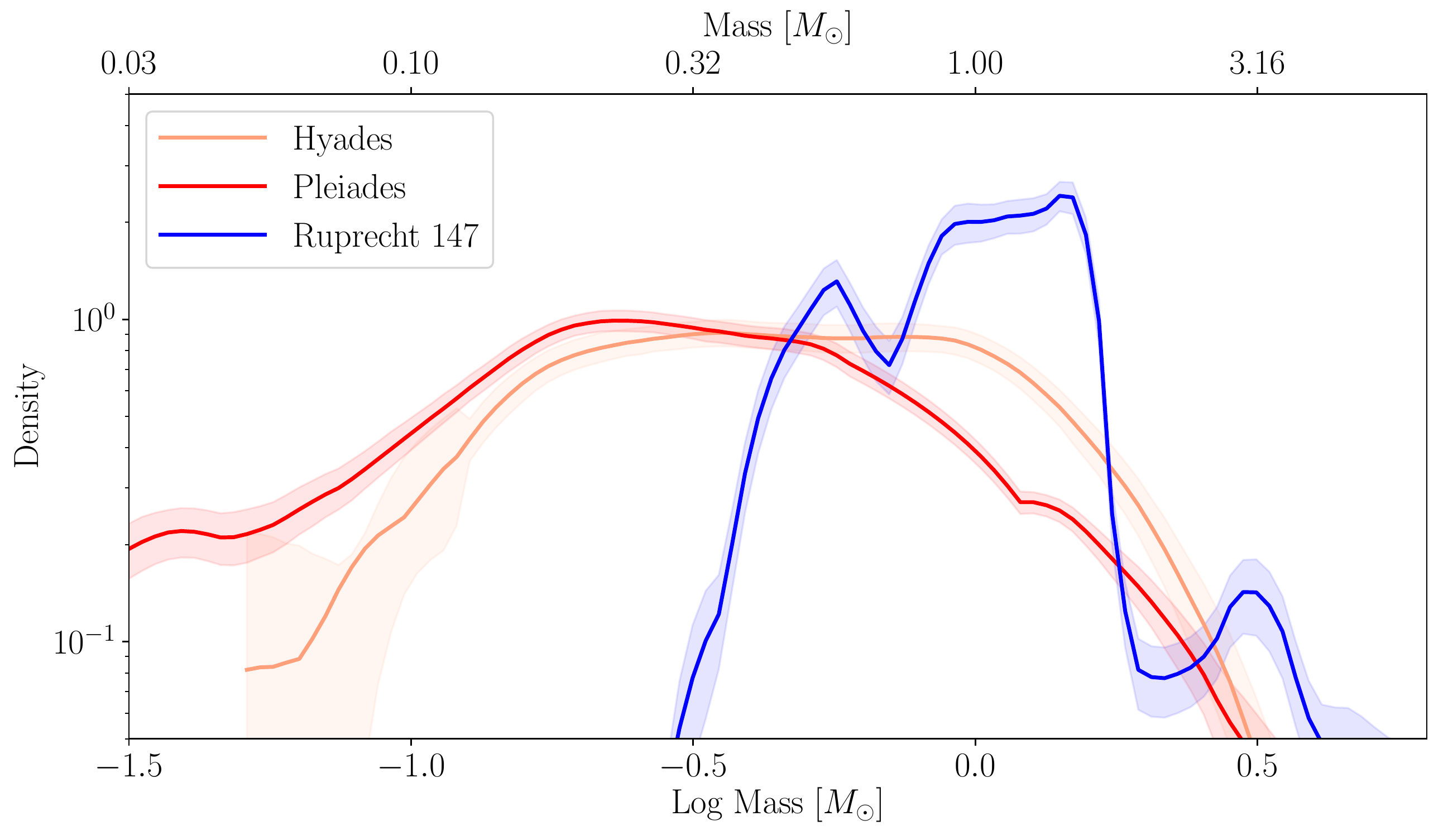}
\caption{Mass distribution of our GDR2 candidate members plus the WDs progenitors as recovered from the COLIBRI isochrone. For comparison, we also show the empirical determination of the mass distribution from: i) the Pleiades cluster from  \citet{2015A&A...575A.120B}, and ii) the Hyades cluster from \citet{ 2013A&A...559A..43G} for masses above 0.15$M_{\odot}$, and \citet{2008A&A...481..661B} for masses below 0.15$M_{\odot}$.}
\label{fig:mass_extended}
\end{center}
\end{figure*}

The extended mass distribution, which incorporate the masses of the WD progenitors, is shown in Table \ref{table:extended_mass} (available at the CDS) and in Figure \ref{fig:mass_extended}. The latter also shows the empirical determination of the Hyades ($\sim$600 Myr) mass distribution from \citet{2008A&A...481..661B}, for masses $M < 0.15M_{\odot}$, and \citet{2013A&A...559A..43G}, for masses $M>0.15M_{\odot}$, and the Pleiades ($\sim$120 Myr)  mass distribution from  \citet{2015A&A...575A.120B}.

From Fig. \ref{fig:mass_extended} we observe the following aspects.
There is a bump at 3.1 $M_{\odot}$ that  is a direct consequence of the plateau in the IFMR \cite[see Fig. 5 of][]{2018ApJ...866...21C}.

When compared to the younger Pleiades or Hyades clusters, Ruprecht 147 lacks many of the low-mass ($<0.4M_{\odot}$) and high-mass ($>1.6M_{\odot}$) stars. 

We hypothesize that the apparent lack of high-mass stars is a consequence of stellar evolution product of the cluster old age. However, this lack can also originate from: i) WDs not recovered by our methodology, ii) other type of evolved stars not included in our analysis, and iii) an incorrect transformation of the final to initial mass. The latter will imply that  IFMR of \citet{2018ApJ...866...21C} is not the appropriate transformation to obtain the initial mass of our 15 WD candidates. 

On the other hand, the lack of low-mass stars can be a consequence of evaporation due to energy equipartition, or ejection produced by dynamical interactions (e.g. three-body encounters, tidal shocks).

\subsection{Projected spatial distribution}
\label{subsection:PSD}
The spatial distribution of stellar systems is key to diagnose their dynamical stage. In this section we fit a series of parametric models to the projected spatial distribution (i.e. in the plane of the sky, hereafter PSD) of our candidate members. We use the parametric models and methodology of \citet{2018A&A...612A..70O}. Those authors use the \textit{PyAspidistra} code and their Pleiades candidate members to select the best PSD model and infer the posterior distributions of its parameters. Here, we work with the same PSD models: \citet{1987ApJ...323...54E}, hereafter EFF, Generalized Density Profile, hereafter GDP (although \citealt{2010MNRAS.407.2241K} call it Nukker), Generalised King \citep{2018A&A...612A..70O}, hereafter GKing, \citet{1962AJ.....67..471K}, Optimum Generalized King  \citep{2018A&A...612A..70O}, herafter OGKing, and Restricted Generalized Density Profile \citep[see][and references therein]{2018A&A...612A..70O}, herafter RGDP.

In addition to the classical model parameters, the \textit{PyAspidistra} code is able to infer: the coordinates of the cluster center, the cluster ellipticity, and its luminosity segregation (a proxy for the mass segregation). Furthermore, it delivers, as a by-product, the Bayesian evidence of the model. The latter allows a robust model comparison and selection by means of the Bayes factor.

We use the coordinates (J2000.0), \texttt{G} photometric magnitude (the most observed in our candidates), and cluster membership probability of objects in Table \ref{table:members}. After running the \textit{PyAspidistra} code, we obtain the Bayesian evidence of each PSD model together with posterior samples of its parameters. 

Among all analyzed models, those with luminosity segregation are the ones with higher Bayesian evidence, with the EFF rating the highest. We compare the Bayes factors of the latter, with respect to all other models, and according to Jeffrey's scale \citep{Jeffreys61} there is strong evidence favoring the EFF model against the GKing and King ones, and moderate evidence favoring it against the GDP, OGKing and RGDP ones.

Table \ref{table:psd_parameters} shows the median values (with the 16 and 84 percentiles as the uncertainties) of the parameters in all our luminosity segregated models. The interested reader can find the definition of these parameters in \citet{2018A&A...612A..70O}. Suffice here to say that $\alpha_c$, $\delta_c$ are the centre coordinates, $\phi$ is the position angle of the semi-major axis $r_{ca}$ of the core radius\footnote{The core radius is the unit scale of the density profile, thus is completely dependant on the latter.}, and  $r_{cb}$ is the semi-minor axis. For the King family of models, $r_{ta}$ and $r_{tb}$ represent the semi-major and semi-minor axes of the tidal radius, respectively. The exponents of the different models are indicated by $\alpha,\beta$ and $\gamma$. The ellipticity of the core and tidal radii are indicated by $\epsilon_{rc}$ and $\epsilon_{rt}$, respectively. Finally, in our luminosity segregated models the core radius $r_c$, grows linearly with magnitude as

\begin{equation}
    r_c(\theta,\texttt{G})= r_c(\theta) + \kappa\cdot(\texttt{G}-G_{mode}),
\end{equation}
with $\kappa$ as indicated in Table \ref{table:psd_parameters}, and $G_{mode}$ the mode of the \texttt{G} magnitude distribution (see Eq. 13 of \citealp{2018A&A...612A..70O}). 

As can be seen in Table \ref{table:psd_parameters}, the $\kappa$ value is in all models positive and incompatible with zero beyond 1$\sigma$, which we interpret as strong evidence of luminosity segregation.

\begin{table*}[ht!]
  \caption{Parameters of luminosity segregated models.} 
  \label{table:psd_parameters}
\resizebox{\textwidth}{!}{%
  \begin{tabular}{cccccccccccccc}
        \hline
		Model & $\alpha_c$ & $\delta_c$ & $\phi$ & $r_{ca}$ & $r_{cb}$ & $\gamma$ & $\kappa$ & $\epsilon_{rc}$ & $\alpha$ & $\beta$ & $r_{ta}$ & $r_{tb}$ & $\epsilon_{rt}$ \\ 
		& [$^{\circ}$] & [$^{\circ}$] & [rad] & [pc] & [pc]&  & [pc $\cdot$mag$^{-1}$] & &  &  & [pc] & [pc] &  \\
		\hline
		EFF & $289.007^{+0.124}_{-0.075}$ & $-16.358^{+0.098}_{-0.104}$ & $1.40^{+0.63}_{-0.31}$ & $2.96^{+1.82}_{-0.98}$ & $1.58^{+0.59}_{-0.48}$ & $2.83^{+0.34}_{-0.44}$ & $0.168^{+0.084}_{-0.119}$ & $0.45^{+0.22}_{-0.26}$ & -- & -- & -- & -- & -- \\ 
		GDP & $289.028^{+0.126}_{-0.085}$ & $-16.358^{+0.095}_{-0.105}$ & $1.38^{+0.66}_{-0.32}$ & $3.4^{+2.4}_{-1.3}$ & $1.78^{+0.85}_{-0.67}$ & $0.30^{+0.36}_{-0.21}$ & $0.18^{+0.12}_{-0.15}$ & $0.47^{+0.23}_{-0.27}$ & $0.47^{+0.40}_{-0.27}$ & $2.71^{+0.53}_{-0.67}$ & -- & -- & -- \\ 
		GKing & $289.013^{+0.125}_{-0.086}$ & $-16.36\pm 0.10$ & $1.34^{+0.48}_{-0.28}$ & $2.55^{+2.12}_{-0.98}$ & $1.35^{+0.72}_{-0.54}$ & -- & $0.15^{+0.11}_{-0.13}$ & $0.45\pm 0.27$ & $0.58^{+0.54}_{-0.32}$ & $2.27^{+0.53}_{-0.78}$ & $119^{+349}_{-78}$ & $42^{+43}_{-25}$ & $0.62^{+0.26}_{-0.36}$ \\ 
		King & $289.013^{+0.127}_{-0.076}$ & $-16.348^{+0.094}_{-0.107}$ & $1.36^{+0.46}_{-0.29}$ & $1.85^{+1.86}_{-0.58}$ & $0.98^{+0.55}_{-0.25}$ & -- & $0.130^{+0.074}_{-0.067}$ & $0.45^{+0.26}_{-0.27}$ & -- & -- & $82^{+140}_{-35}$ & $34^{+12}_{-14}$ & $0.59^{+0.26}_{-0.33}$ \\ 
		OGKing & $289.008^{+0.131}_{-0.079}$ & $-16.349^{+0.095}_{-0.106}$ & $1.38^{+0.48}_{-0.30}$ & $1.97^{+1.84}_{-0.60}$ & $1.06^{+0.54}_{-0.27}$ & -- & $0.128^{+0.073}_{-0.100}$ & $0.44^{+0.25}_{-0.27}$ & -- & -- & $88^{+171}_{-40}$ & $36\pm 14$ & $0.59^{+0.27}_{-0.34}$ \\ 
		RGDP & $289.013^{+0.134}_{-0.083}$ & $-16.36^{+0.10}_{-0.11}$ & $1.40^{+0.65}_{-0.35}$ & $3.1^{+2.0}_{-1.2}$ & $1.56^{+0.72}_{-0.61}$ & -- & $0.16^{+0.11}_{-0.14}$ & $0.47^{+0.23}_{-0.26}$ & $0.56^{+0.39}_{-0.31}$ & $2.79^{+0.58}_{-0.63}$ & -- & -- & -- \\ 
		\hline
    \end{tabular}%
}
\end{table*}

\begin{figure*}
\begin{center}
\includegraphics[width=0.6\textwidth]{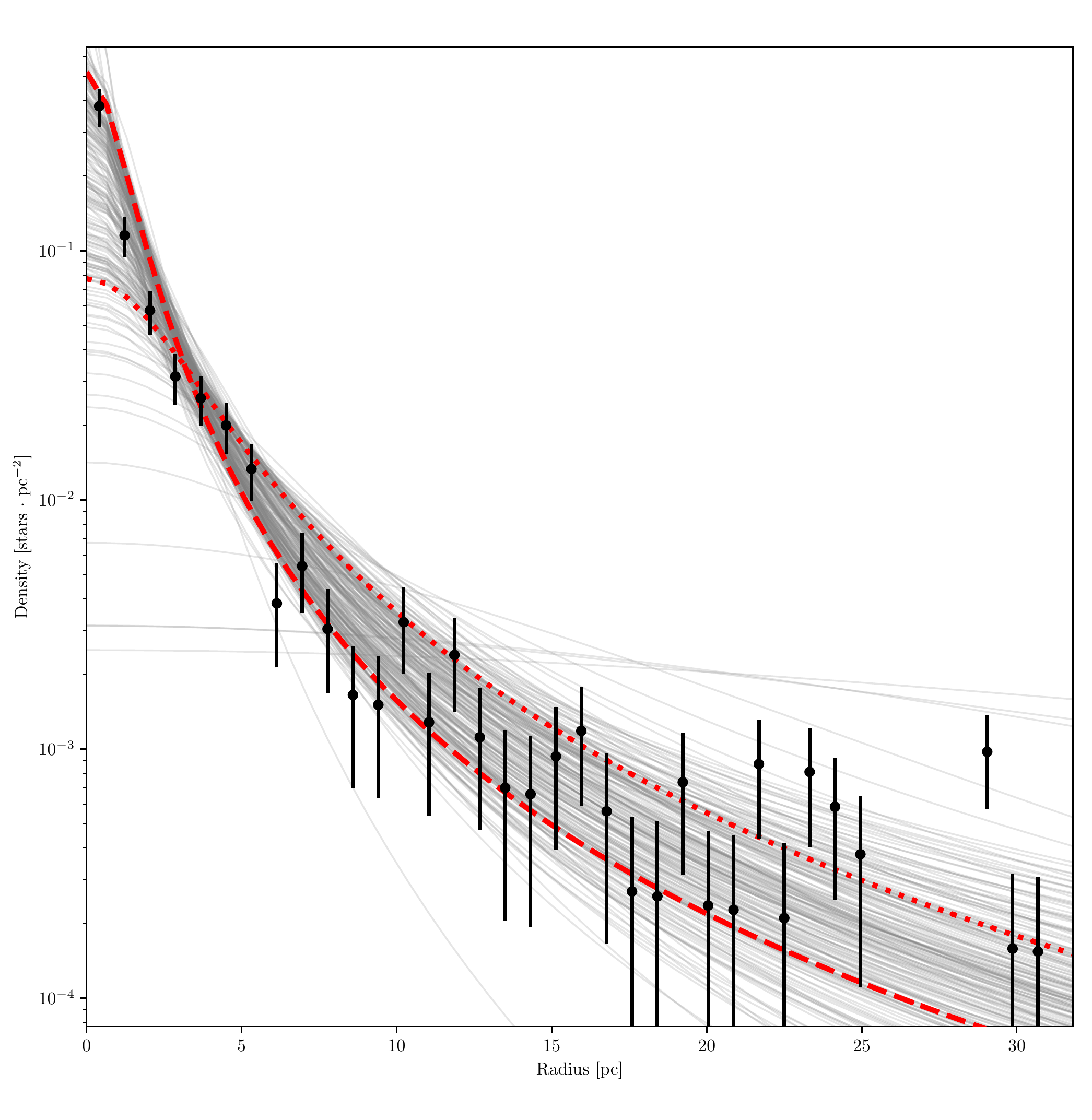}
\includegraphics[width=0.6\textwidth]{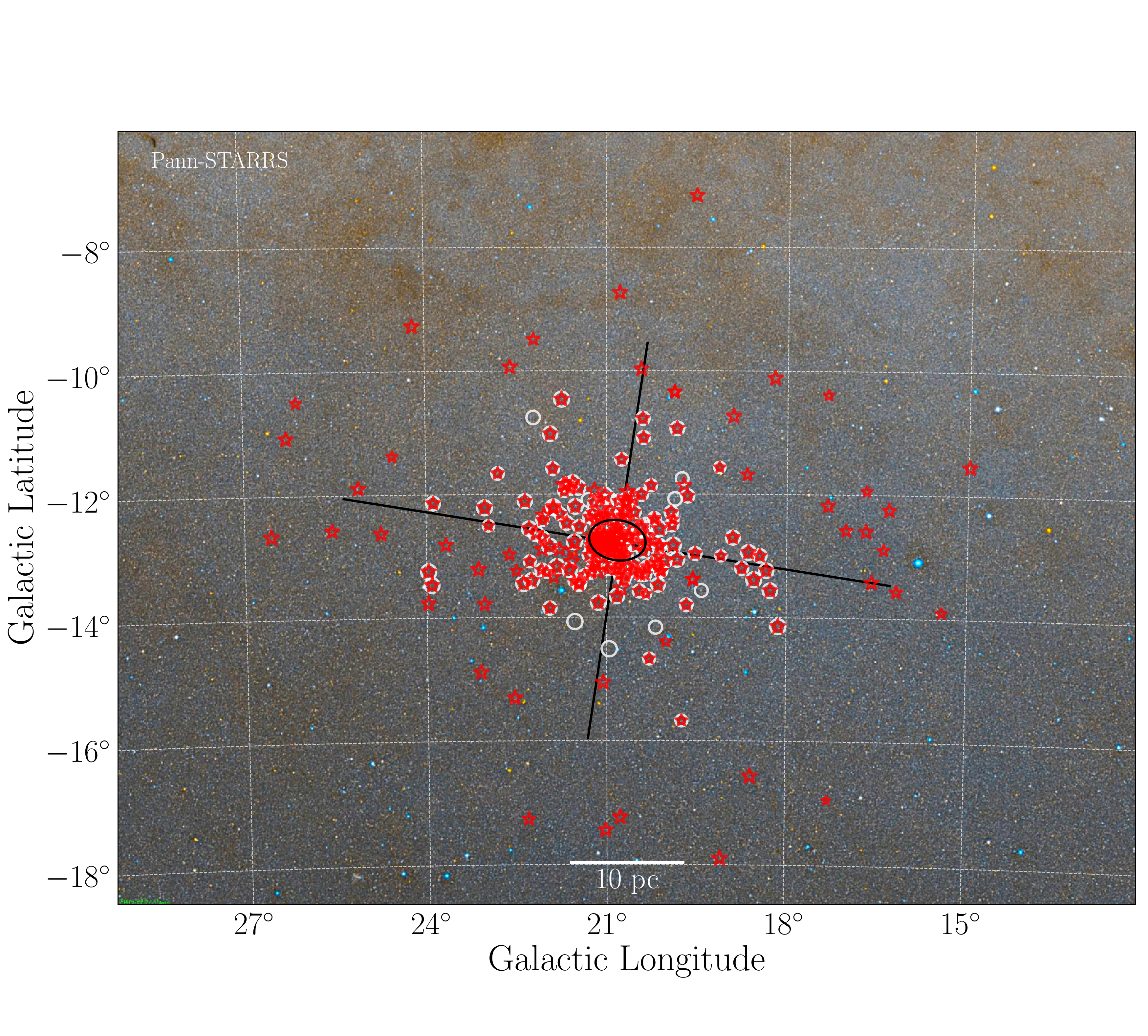}
\caption{The EFF Segregated model. Top panel: Density of candidate members (black dots with Poisson uncertainties) together with: the best-fitting model (computed from the semi-major and semi-minor axes of the core radius and plotted in dotted and dashed red lines, respectively) and samples from the posterior distribution (grey lines). Bottom panel: Galactic coordinates of our candidate members (red stars), and those of \citet[][white circles]{Cantat-Gaudin2018}. Also plotted is the core radius (black ellipse) and two perpendicular (black) lines extending ten core radius units in the directions of the semi-major and semi-minor axes.}
\label{fig:PSD}
\end{center}
\end{figure*}

The top panel of Figure \ref{fig:PSD} shows the density of candidate members as a function of radial distance and the best-fitting profile of the luminosity segregated EFF model, together with samples from the posterior distribution of its parameters. In addition, the bottom panel of Fig. \ref{fig:PSD} shows the galactic coordinates of our candidate members, together with an ellipse depicting the core radius. As observed in this panel, the cluster candidate members are homogeneously distributed in position angle. 

Furthermore, the family of King profiles allow us to estimate the total number of cluster stars contained within the tidal radius \citep[see Eq. 18 of][]{1962AJ.....67..471K}. The GKing, King and OGKing predict $278_{-17}^{+48}$, $276_{-16}^{+59}$,$278_{-17}^{+60}$ cluster stars, respectively, which means that we are still missing $\sim$5\% of the cluster population. This percentage of missing members is the same predicted by the TPR$\sim$95\% of our membership selection (see Appendix \ref{appendix:sensitivity}). Thus, we expect only a negligible fraction of cluster members beyond the limits imposed in Assumption \ref{ass:6deg}. 

Our analysis of the cluster PSD shows the following important elements:

\begin{itemize}
    \item The best profile is the EFF.
    \item There is strong evidence of luminosity (mass) segregation.
    \item The cluster is highly elliptic ($\epsilon \gtrsim 0.4)$.
    \item The semi-major axis of the cluster is in close alignment with the galactic plane.
\end{itemize}

\section{Summary and conclusions}
\label{section:summary}

In this Section, we briefly summarize our results, discuss the consequences of our assumptions, and present our conclusions.

\subsection{Summary}

Our combined analysis over the \textit{Gaia} DR2 and DANCe data sets enable us to obtain cluster candidate members with masses heavier than $\sim$0.1$M_{\odot}$. This was possible only thanks to the combination of the exquisite \textit{Gaia} astrometry with the deep and extend DANCe photometry.

The improvements to \citet{2014A&A...563A..45S} methodology allowed us to incorporate parallaxes and the full covariance matrix of the uncertainties in the cluster model construction. In addition, by modelling the WDs photometry, we are now able to incorporate these objects into the cluster mass distribution.

On the sensitivity tests performed over the  GDR2, DANCe and DANCe Central data sets (see Appendix \ref{appendix:sensitivity}) we find that the results of our improved methodology are still sensitive to the $p_{in}$ threshold, but only for certain interval of values. We objectively set the  $p_{in}$ as the minimum of the plateau where the results are insensitive to the specific value of $p_{in}$ used.

We found 259 candidate members (see Table \ref{table:members}), 58 of them not previously known. There is a general agreement between both our lists and those from the literature, and the general properties derived from them. 

The empirical isochrone given by our candidates is in excellent agreement with theoretical ones, with the COLIBRI one particularly.

The inferred the $A_V$ extinction peaks at $A_V=0.36$ mag, with a large dispersion (from 0 to 1.4 mag) which is also present in the $A_G$ values reported by \textit{Gaia}. We interpret it as evidence of differential extinction, as originally proposed by \citet{2013AJ....145..134C}

We observe an increase of low-luminosity objects ($\log L/L_{\odot} < -1.7$) at the outer parts of the cluster that we interpret as evidence of luminosity segregation. The latter is confirmed by our analysis of the cluster PSD. 

The Bayesian evidence suggest that the best model for the PSD of the cluster is the luminosity segregated EFF. In addition, the cluster PSD is highly elliptical ($\epsilon\sim 0.45$).

\subsection{Discussion of assumptions}

We have assumed that the cluster members are contained within a 6$^\circ$ radius area (Assumption \ref{ass:6deg}). However, from our analysis of the PSD we find that we are typically missing 5\% of them. The latter is similar to that predicted by our membership selection process (measured in the analyzed data sets). Thus, we expect a negligible fraction of cluster members beyond the assumed 6$\degr$ radius.

We have assumed that the missing values are uniformly distributed (Assumption \ref{ass:MCAR}). However, missing values in the photometry are more frequent at faint magnitudes, near the detection threshold. Because of this, the density of sources in our field model can be underestimated at this faint magnitudes, which might result in unexpected contaminants.

We have assumed that the cluster distribution originates only from single, binaries and WD stars (Assumption \ref{ass:populations}). Due to this assumption, our methodology still misses multiple systems, and stars in other evolved stages. For example, \citet{2016PhDT.......246C} reports seven Blue Stragglers as candidate members of Ruprecht 147 (see Section 3.4.5 of the mentioned work). From these, our methodology recovers only three (CWW 62, 150 and 152). Future steps will be taken to incorporate Blue Stragglers and other stellar populations into our cluster model.

We have assumed that the cluster age is 2.5 Gyr (Assumption \ref{ass:cluster_age}). However, it should be revisited in the light of the WDs candidates reported here. The latter can in principle, be used to provide and independent estimate of the cluster age.  

We have assumed that the theoretical isochrones provide the true mass-luminosity relation (Assumption \ref{ass:isochrones}). However, the Wielen and Kroupa dips in the mass distribution could be artifacts of an incorrect mass-luminosity relation.

We have assumed that the mass of the WD progenitor can be recovered from the WD IFMR of \citet{2018ApJ...866...21C}. However, this relation might not be the appropriate one. 

\subsection{Conclusions}
\label{subsection:conclusions}

The detailed compilation and analysis of astrometric and photometric data provided by the COSMIC-DANCe methodology, combined with the exquisite \textit{Gaia} data, and our comprehensive improved methodology, enable us to derive luminosity and mass distribution with a superb degree of detail. 
We recover 259 candidate members, which represents an improvement of 30\% with respect to previous studies.

The clusters is mass segregated, high elliptical, and lacks several low-mass stars. All these indicators can be used in the future to reconstruct its dynamical history.

Nevertheless, there are still aspects that remain open:
\begin{itemize}
\item Binaries (and therefore multiple systems) may produce spurious astrometric and photometric results that negatively impact the recovering of possible cluster members.
\item The spectroscopic and eclipsing binaries of Ruprecht 147 \citep[e.g.][]{2018ApJ...866...67T, 2016PhDT.......246C} can be used to further constrain the mass-luminosity relation, which will shed light in the understanding of the Wielen dip.

\item The presence of the Kroupa dip in Ruprecht 147 is still uncertain. Better statistics are still needed to confirm or discard its presence. 

\item A detailed analysis of the fraction of binaries and hierarchical systems can help to reconstruct the initial mass distribution, which will be an important step towards the understanding the universality of the initial mass distribution.
\end{itemize}
\begin{acknowledgements}
We acknowledge Jason Lee Curtis for his kind comments, which considerably improved the quality of this work.

This research has received funding from the European Research 
Council (ERC) under the European Union’s Horizon 2020 research and innovation programme (grant agreement No 682903, P.I. H. Bouy), and from the French State in the framework of the ”Investments for the future” Program, IdEx Bordeaux, reference ANR-10-IDEX-03-02 .

This research draws upon data distributed by the NOAO Science Archive. NOAO is operated by the Association of Universities for Research in Astronomy (AURA) under cooperative agreement with the National Science Foundation. 

This publication makes use of data products from the Two Micron All Sky Survey, which is a joint project of the University of Massachusetts and the Infrared Processing and Analysis Center/California Institute of Technology, funded by the National Aeronautics and Space Administration and the National Science Foundation.  

This work has made use of data from the European Space Agency (ESA)
mission {\it Gaia} (\url{http://www.cosmos.esa.int/gaia}), processed by the {\it Gaia} Data Processing and Analysis Consortium (DPAC,
\url{http://www.cosmos.esa.int/web/gaia/dpac/consortium}). Funding
for the DPAC has been provided by national institutions, in particular the institutions participating in the {\it Gaia} Multilateral Agreement.

This research has made use of the VizieR and Aladin images and catalog access tools and of the SIMBAD database, operated at the CDS, Strasbourg, France. 
This publication makes use of data products from the Wide-field Infrared Survey Explorer, which is a joint project of the University of California, Los Angeles, and the Jet Propulsion Laboratory/California Institute of Technology, funded by the National Aeronautics and Space Administration. 

Some/all of the data presented in this paper were obtained from the Mikulski Archive for Space Telescopes (MAST). STScI is operated by the Association of Universities for Research in Astronomy, Inc., under NASA contract NAS5-26555.

\end{acknowledgements}


\begin{table}
\centering
\caption{Sources in the GDR2 catalogue.}
\label{table:gaia_data_and_probabilities}
\end{table}

\begin{table}
\centering
\caption{Sources in the DANCe catalogue.}
\label{table:dance_data_and_probabilities}
\end{table}

\begin{table}
\centering
\caption{Sources in the DANCe Central catalogue.}
\label{table:dance-central_data_and_probabilities}
\end{table}

\begin{table}
\centering
\caption{Ruprecht 147 Candidate members.}
\label{table:members}
\end{table}

\begin{table}
\centering
\caption{Empirical isochrone in the \textit{Gaia} photometry.}
\label{table:isochrone_gaia}
\end{table}

\begin{table}
\centering
\caption{Empirical isochrone in the DANCe photometry.}
\label{table:isochrone_dance}
\end{table}

\begin{table}
\centering
\caption{Extended mass distribution.}
\label{table:extended_mass}
\end{table}


\begin{appendix}

\section{Sensitivity and quality of the classifier}
\label{appendix:sensitivity}

In this Appendix, we analyze the sensitivity of our methodology to three elements, the representation space, the initial list of members, and the $p_{in}$ value used. In addition, we also report the quality of the classifier as function of magnitude and optimum probability threshold.

\subsection{Sensitivity to representation space}
We analyze the sensitivity of our methodology to the representation space in the DANCe Central data set. We chose the latter because it does not have the highly discriminant parallax feature, and it is also the one that reaches fainter photometric magnitudes. 

We run our methodology on the representation spaces $RS_1$ and $RS_2$. Both were chosen based on the importance of each feature, as given by the random forest classifier (see Section \ref{section:membership_selection}). However, the $RS_2$ features were selected after training the random forest classifier on the candidate members found on $RS_1$. The last two blocks of Table \ref{table:TPR_CR} report the number of recovered members, TPR and CR as function of the $p_{in}$ obtained from the $RS_1$ and $RS_2$ in the DANCe Central data set. We observe that at each $p_{in}$, the values reported by the two representation spaces are compatible with each other within their uncertainties (using Poisson uncertainties for the number of members). Therefore, we conclude that our methodology is statistically insensitive to the representation space, provided that is photometric features are chosen from within those with higher importance.

\subsection{Sensitivity to initial list of members}

\begin{figure}[htp]
    \centering
    \includegraphics[width=\columnwidth]{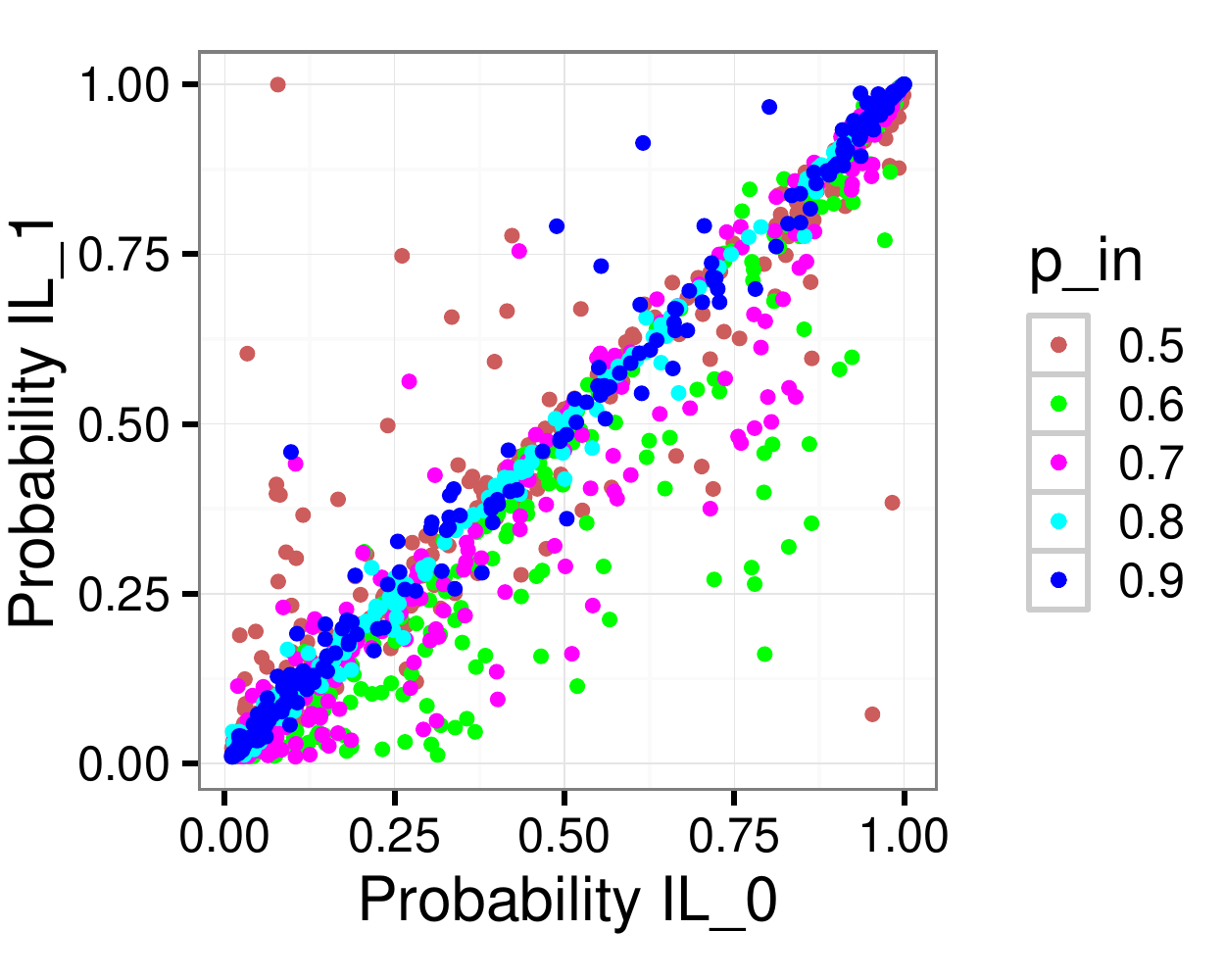}
    \caption{Comparison of the recovered membership probabilities from the two initial lists of members at different $p_{in}$ values (colour coded).}
    \label{fig:Cantat_vs_Central}
\end{figure}

We analyze the sensitivity of our methodology to the initial list of candidate members in the DANCe data set. We use as initial lists the candidate members described in Section \ref{subsection:inital_list}, hereafter $IL_0$, and those found after applying our methodology to the DANCe Central data set, hereafter $IL_1$. These two list have different limiting magnitudes. While the first one, which is that of \citet{Cantat-Gaudin2018} plus the WDs, reaches only \texttt{G} $\sim$ 18, the second one, obtained in the DANCe Central data set reaches the faintest magnitudes (\texttt{r}$\sim20$). Figure \ref{fig:Cantat_vs_Central} compares,  for different $p_{in}$ values, the membership probabilities recovered using the two initial lists. As can be seen, there are only a few objects with probabilities departing from the identity relation, and for lower $p_{in}$ values particularly. Nevertheless, the number of these objects is negligible compared to the size of the data set ($\sim$7 millions). Therefore, we conclude that our methodology is insensitive to the initial list of members.

\subsection{Sensitivity to $p_{in}$ values}

The $p_{in}$ value establishes the probability threshold to select objects as part of the cluster model (see Sect. \ref{section:membership_selection}), thus, we expect the results of our methodology to be sensitive to it. As described in Sect. \ref{subsection:iterative_classification}, we run our methodology in our three data sets using $p_{in}$ values in the range 0.5 to 0.9 in steps of 0.1. 
In Table \ref{table:TPR_CR}, we report the optimum probability threshold ($p_{opt}$), number of candidate members, TPR and CR recovered for each data set and $p_{in}$ value used.

\begin{figure}[ht!]
\begin{center}
\includegraphics[width =\columnwidth]{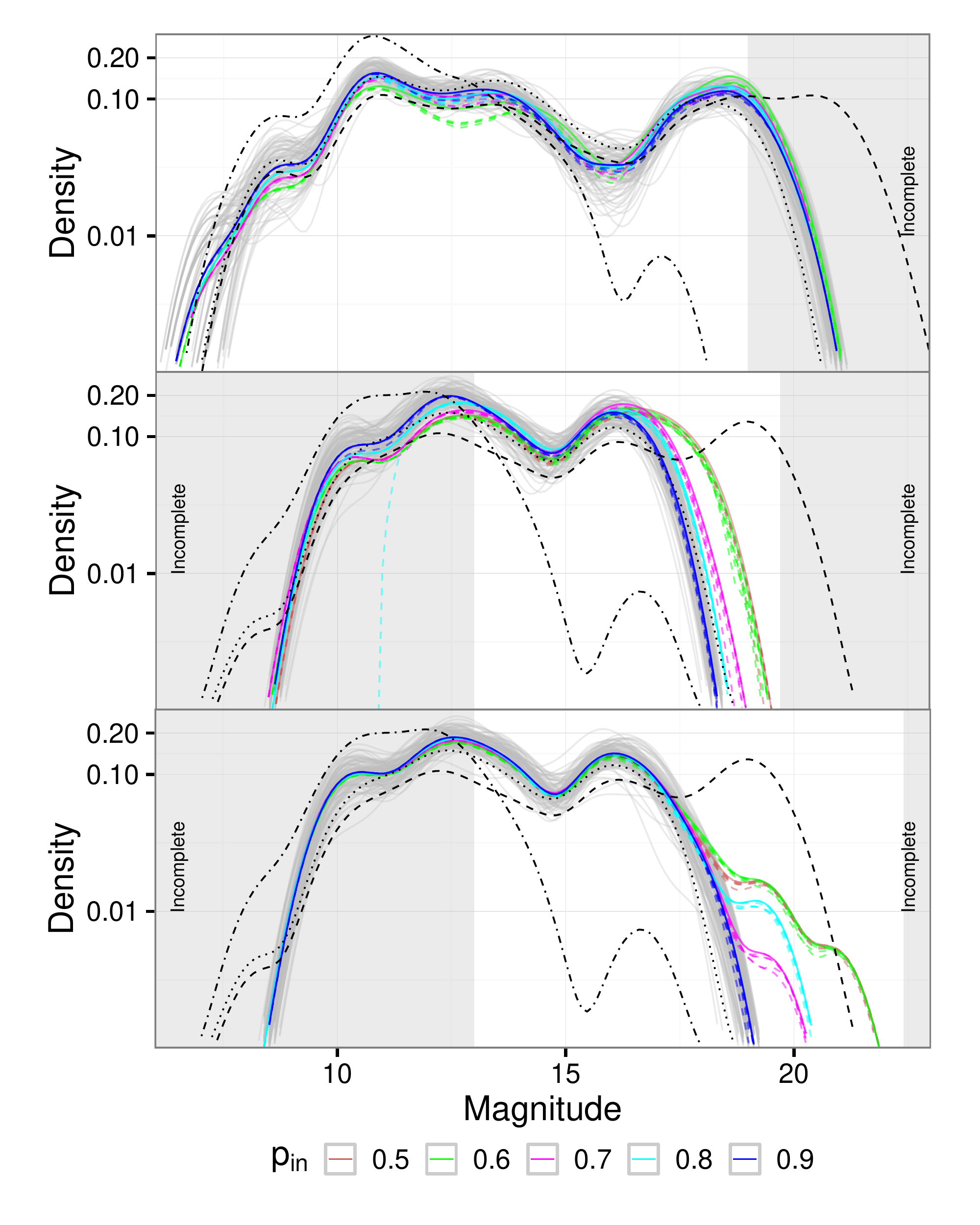}
\caption{Magnitude distributions computed from the candidate members recovered by the GDR2 (top panel at \texttt{BP} band), DANCe (middle panel at \texttt{i} band) and DANCe Central (bottom panel at \texttt{i} band) classifiers at different $p_{in}$ values (colour code).
Also shown are the magnitude distributions: i) of our candidate members corrected by the PPV (dashed lines),  ii) of bootstrap realizations from the candidate members at the higher $p_{in}$ value (grey lines), and iii) of the candidate members from \citet{2013AJ....145..134C}, \citet{Babusiaux2018} and \citet{Cantat-Gaudin2018}, in dot-dashed, dashed, and dotted lines, respectively.}
\label{fig:magnitude_distributions}
\end{center}
\end{figure}

In addition, Fig. \ref{fig:magnitude_distributions} shows the magnitude distributions of the candidate members recovered in our data sets and for all the $p_{in}$ values used. This Fig. also show: i) one hundred bootstrap realizations of the candidate members resulting from $p_{in}=0.9$, and ii) the magnitude distributions corrected from the PPV (see next section).

As can be observed from Table \ref{table:TPR_CR} and Fig. \ref{fig:magnitude_distributions}, the TPR, CR, number of recovered members and magnitude distributions are sensitive to the $p_{in}$ value, and cannot be reconciled by their uncertainties. However, we also observe that there are ranges of $p_{in}$ values that return consistent results. Thus, we now proceed to identify these ranges, and to establish their minimum $p_{in}$ value. 

\begin{table*}[ht!]
\caption{Quality of the GDR2, DANCe and DANCe Central classifiers at each $p_{in}$ value used.}
\label{table:TPR_CR}
\resizebox{\textwidth}{!}{%
\begin{tabular}{|c|cccc|cccc|cccc|cccc|}
\hline
$p_{in}$& \multicolumn{4}{|c|}{GDR2 ($RS_0$)} & \multicolumn{4}{|c|}{DANCe ($RS_1$)}   & \multicolumn{4}{|c|}{DANCe Central ($RS_1$)} & \multicolumn{4}{|c|}{DANCe Central ($RS_2$)} \\
\hline
        & $p_{opt}$ & Mem. & TPR & CR & $p_{opt}$ & Mem. & TPR & CR& $p_{opt}$ & Mem. & TPR & CR & $p_{opt}$ & Mem. & TPR & CR \\
\hline
\hline
0.5     &    --- & --  &                -- & --                 & 0.778 & 159  & $0.926 \pm 0.008$ & $0.061 \pm 0.008$ & 0.818 & 128 & $0.983 \pm 0.005$ & $0.022 \pm 0.004$ & 0.848 & 132 & $0.971 \pm 0.008$ & $0.034 \pm 0.005$ \\ 
0.6     &  0.828 & 299 & $0.911 \pm 0.003$ & $0.122 \pm 0.002$  & 0.747 & 150  & $0.931 \pm 0.006$ & $0.062 \pm 0.007$ & 0.737 & 142 & $0.983 \pm 0.009$ & $0.021 \pm 0.007$ & 0.778 & 136 & $0.976 \pm 0.003$ & $0.040 \pm 0.007$ \\ 
0.7     &  0.859 & 219 & $0.950 \pm 0.007$ & $0.072 \pm 0.010$  & 0.697 & 140  & $0.944 \pm 0.011$ & $0.050 \pm 0.009$ & 0.717 & 137 & $0.971 \pm 0.003$ & $0.022 \pm 0.004$ & 0.869 & 124 & $0.967 \pm 0.005$ & $0.023 \pm 0.004$ \\ 
0.8     &  0.869 & 198 & $0.956 \pm 0.006$ & $0.050 \pm 0.007$  & 0.727 & 117  & $0.951 \pm 0.005$ & $0.039 \pm 0.007$ & 0.798 & 125 & $0.972 \pm 0.012$ & $0.022 \pm 0.005$ & 0.808 & 128 & $0.979 \pm 0.005$ & $0.029 \pm 0.008$ \\ 
0.9     &  0.848 & 175 & $0.957 \pm 0.007$ & $0.045 \pm 0.006$  & 0.768 &  96  & $0.949 \pm 0.012$ & $0.032 \pm 0.007$ & 0.798 & 115 & $0.975 \pm 0.006$ & $0.018 \pm 0.007$ &  0.859 & 118 & $0.975 \pm 0.002$ & $0.026 \pm 0.004$ \\ 
\hline
\end{tabular}%
}
\tablefoot{The GDR2 classifier did not converged at $p_{in}=0.5$.}
\end{table*}

In the GDR2 data set we chose to work with the $p_{in}=0.7$ because of the following reasons.
\begin{itemize}
    \item It results in CR, TPR and $p_{opt}$ similar to those of higher values. 
    \item It recovers the candidate members of higher $p_{in}$ values.
    \item It produces magnitude distributions ( both the corrected and uncorrected by the PPV) that are compatible, within the uncertainties, with those of higher $p_{in}$ values (see Fig. \ref{fig:magnitude_distributions}).
\end{itemize}

The results of the DANCe data set seem to be more sensitive to the $p_{in}$ value than those of the GDR2 data set, at the faint end particularly (\texttt{i}>18 mag). In this data set we chose to work with the $p_{in}=0.7$ because:
\begin{itemize}
    \item It recovers the candidate members of higher $p_{in}$ values.
    \item It produces magnitude distributions (both the corrected and uncorrected by the PPV) that are compatible, within the uncertainties, with those of higher $p_{in}$ values (see Fig. \ref{fig:magnitude_distributions}).
\end{itemize}

The results of the DANCe Central data set are virtually insensitive to the $p_{in}$ values. However, the magnitude distributions vary with $p_{in}$ at the faint end ( \texttt{i}> 19~mag),  these variations are negligible though. The latter are produced by four candidate members (see Fig. \ref{fig:CMD}) that are recovered by $p_{in}=\{0.5,0.6\}$ but not by $p_{in}=0.9$. These four objects are of particular interest since they lay near and beyond \textit{Gaia} completeness limit. Thus, we chose to work with the $p_{in}=0.6$ because:

\begin{itemize}
    \item It recovers the candidate members of higher $p_{in}$ values.
    \item Its magnitude distributions (both the corrected and uncorrected by the PPV) at \texttt{i}< 19~mag are compatible, within the uncertainties, with those of higher $p_{in}$ values (see Fig. \ref{fig:magnitude_distributions}).
    \item It recovers faint candidate members beyond \textit{Gaia} completeness limit.
\end{itemize}

\subsection{Quality of the classifier}

In this Section, we analyze the quality of the classifiers at their chosen $p_{in}$ values (as established above) and for all probability thresholds and magnitude ranges. 

\begin{figure*}
\begin{center}
\includegraphics[width = 0.32\textwidth,page=3]{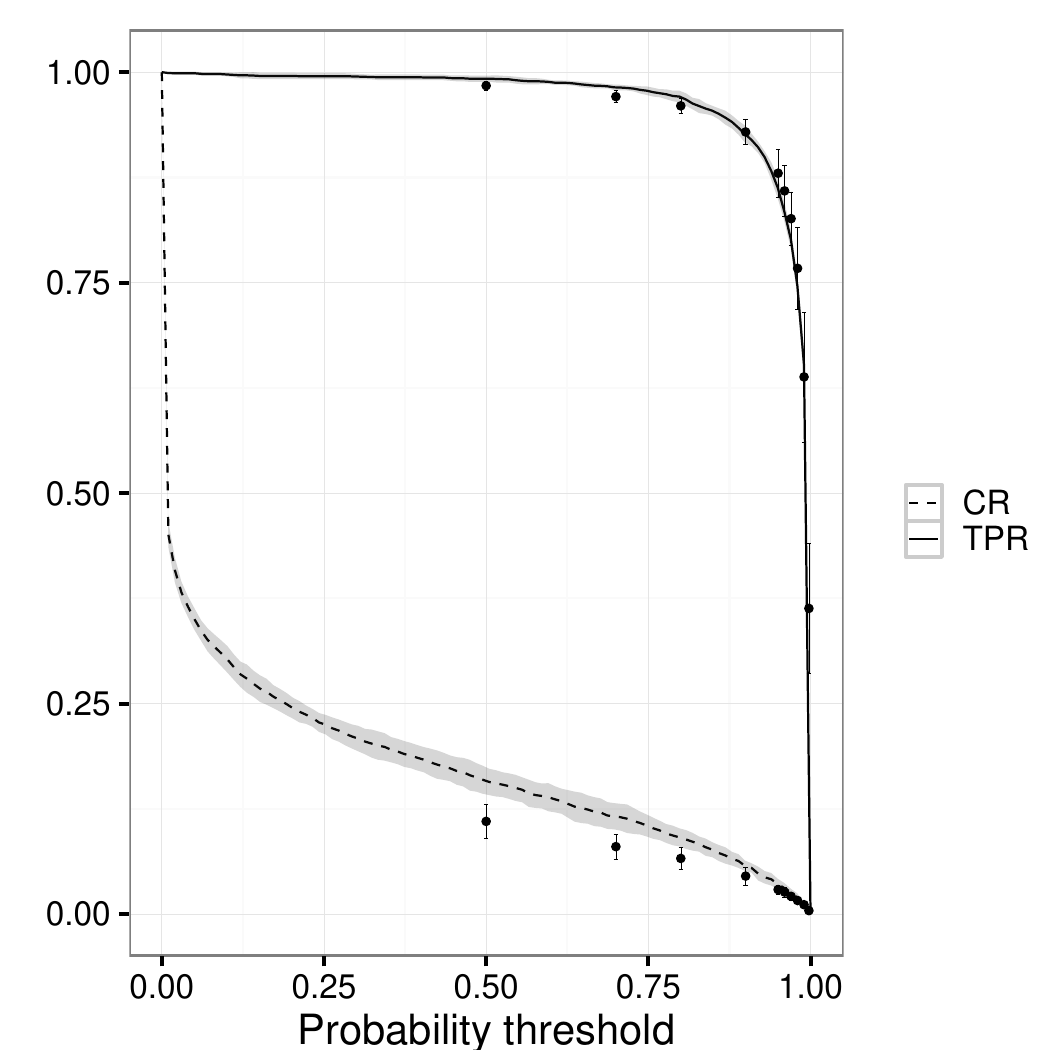}
\includegraphics[width = 0.32\textwidth,page=3]{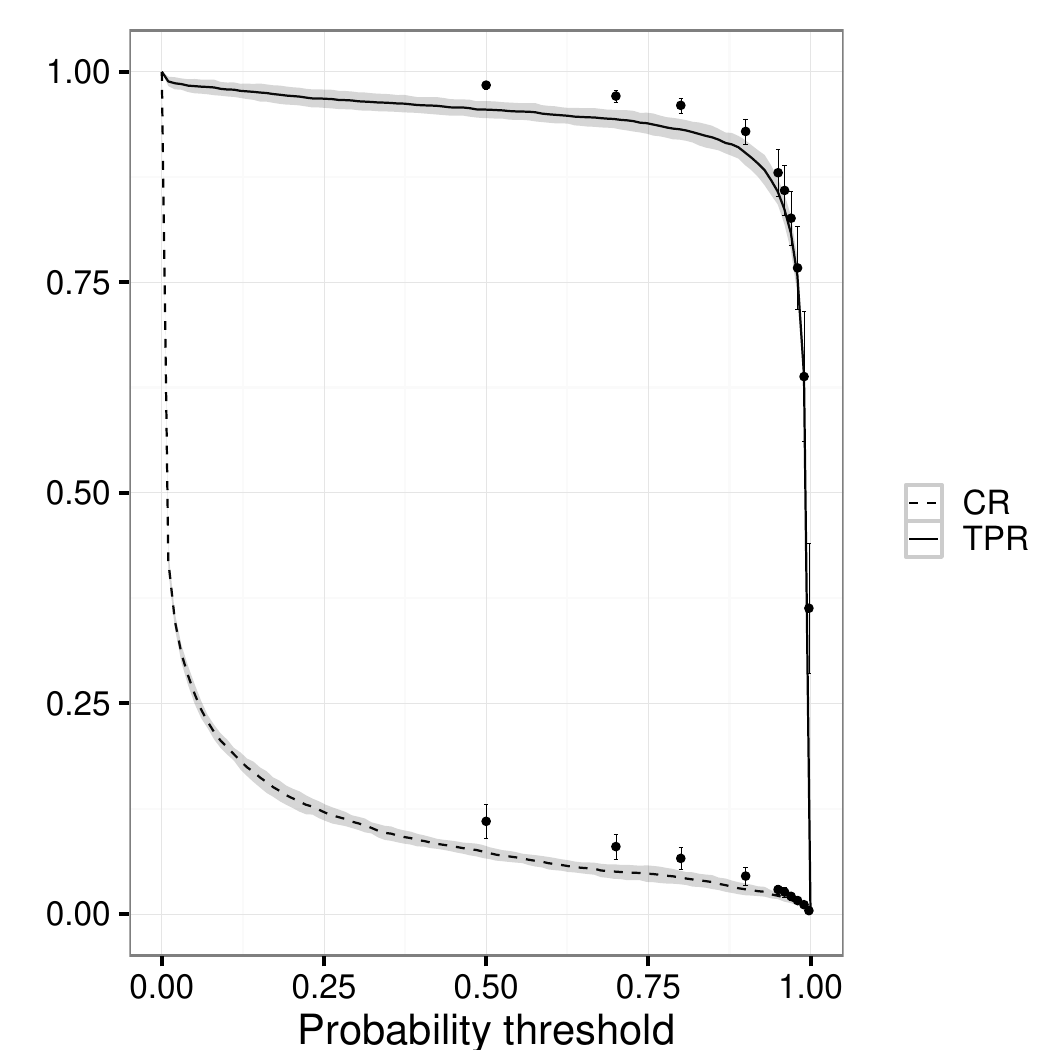}
\includegraphics[width = 0.32\textwidth,page=3]{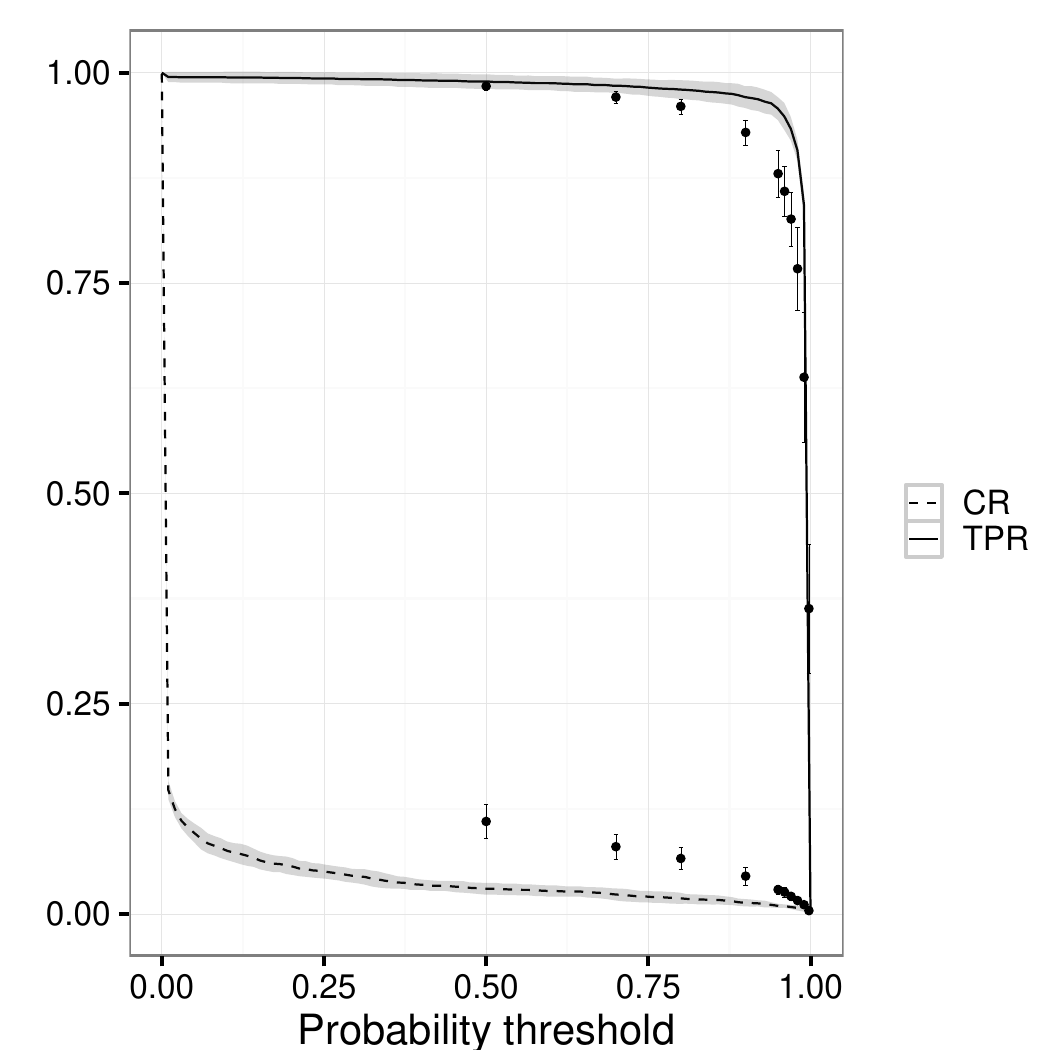}
\caption{True Positive Rate and Contamination Rate as functions of the classification threshold for the GDR2 ($p_{in}=0.7$, left panel), DANCe ($p_{in}=0.7$, middle panel) and DANCe Central ($p_{in}=0.6$, right panel) classifiers. Gray areas show the standard deviation computed from the five synthetic data sets. The dots show the TPR and CR reported by \citet{2014A&A...563A..45S} for their Pleiades classifier.}
\label{fig:TPR_CR}
\end{center}
\end{figure*}

In Fig. \ref{fig:TPR_CR} we show the TPR and CR of our classifiers for all probability thresholds. For comparison, the figure also shows the TPR and CR found by \citet{2014A&A...563A..45S}. From the previous figure we observe the following points. 
First, as expected the quality of the classifier improves proportionally to the ratio of cluster members to field contaminants. In spite of that, the GDR2 classifier has a higher TPR that the DANCe one due to the use of the highly discriminant parallax feature. 
Second, all classifiers have a range of probability thresholds that render similar TPR and CR. Thus, there is a range of values around the $p_{opt}$ that result in similarly good TPR and CR. 
Fourth, our methodology renders classifiers with similarly good TPR and CR as that of \citet{2014A&A...563A..45S}.

Now, we examine the quality of the classifiers as a function of the photometric magnitude. The latter must be present in the representation space in order to generate appropriate synthetic samples for its analysis. Then, we create magnitude bins for these samples, and in each of them we measure the Positive Predictive Value (PPV herafter) at the $p_{opt}$, with the PPV defined as

\begin{equation}
    PPV=\frac{TP}{TP+FP}.
\end{equation}

Finally, we recover the number of TPs multiplying the magnitude distribution by the PPV measured at that magnitude value (we do this by interpolating from the magnitude bins). 
The magnitude distributions corrected by the PPV are shown as dashed lines in Fig. \ref{fig:magnitude_distributions}. As can be seen, the corrected distributions at the chosen $p_{in}$ values are indistinguishable from the uncorrected ones, thus probing that the number of rejected TPs is negligible.

\end{appendix}

\bibliographystyle{aa} 
\bibliography{mybiblio.bib}

\end{document}